\documentclass[11pt]{article}

\usepackage[latin1]{inputenc}
\usepackage{hyperref}
\usepackage{latexsym}
\usepackage{physics}
\usepackage{cancel}
\usepackage{braket}
\usepackage{times}
\usepackage{mathtools}
\usepackage{xcolor}
\usepackage{amsmath}
\usepackage{bbm}
\usepackage{yhmath}
\usepackage{amsthm}
\usepackage{leftidx}
\usepackage{amsfonts}
\usepackage{braket}
\usepackage{slashed}

\newenvironment{sciabstract}{%
\begin{quote} }
{\end{quote}}

\newcounter{lastnote}

\title{\huge{String-inspired Methods and the Worldline Formalism in Curved Space}}

\author
{Olindo Corradini,$^{a,b,c}$\footnote{E-mail: olindo.corradini@unimore.it}$\;$   Maurizio Muratori$^{a}$\footnote{E-mail: 166462@studenti.unimore.it} \\
\\
\normalsize{$^{a}$Dipartimento di Scienze Fisiche, Informatiche e Matematiche, Universit\`a di}\\\normalsize{Modena e Reggio Emilia, via Campi 213/A, I-41125 Modena, Italy}\\[2mm]
\normalsize{$^{b}$INFN, Sezione di Bologna, via Irnerio 46, I-40126 Bologna, Italy}\\[2mm]
\normalsize{$^{c}$Max-Planck-Institut f\"ur Gravitationsphysik, Albert-Einstein-Institut}\\ \normalsize{Am M\"uhlenberg 1, 14476 Golm, Germany}
}

\date{}

\newcommand*{\defeq}{\mathrel{\vcenter{\baselineskip0.5ex \lineskiplimit0pt
                     \hbox{\scriptsize.}\hbox{\scriptsize.}}}%
                     =}
\begin{document} 


\baselineskip15pt


\maketitle


\begin{sciabstract}
\begin{center}\bf{Abstract}\end{center}
The worldline approach to Quantum Field Theory (QFT) allows to efficiently compute several quantities, such as one-loop effective actions, scattering amplitudes and anomalies, which are linked to particle path integrals on the circle. A helpful tool in the worldline formalism on the circle, are string-inspired (SI) Feynman rules, which correspond to a specific way of factoring out a zero mode. In flat space this is known to generate no difficulties. In curved space, it was shown how to correctly achieve the zero mode factorization by applying BRST techniques to fix a shift symmetry. Using special coordinate systems, such as Riemann Normal Coordinates, implies the appearance of a non-linear map---originally introduced by Friedan---which must be taken care of in order to obtain the correct results. In particular, employing SI Feynman rules, the map introduces further interactions in the worldline path integrals. In the present paper, we compute in closed form Friedan's map for RNC coordinates in maximally symmetric spaces, and test the path integral model by computing trace anomalies. Our findings match known results. 

\vspace*{\fill}
\end{sciabstract}


\section{Introduction}
\label{sec:intro}
In the worldline approach to Quantum Field Theory (QFT), particle path integrals are used as a versatile computational tool.  
The method was introduced by Feynman who, already in the 1950, proposed a particle model representation for the dressed scalar propagator in scalar Quantum Electrodynamics~\cite{Feynman:1950ir}. However, it was only in the late 80's that the method started to be taken seriously as an alternative approach to conventional second-quantized methods. Initially it was used as a tool to compute chiral anomalies~\cite{AlvarezGaume:1983at, AlvarezGaume:1983ig, Friedan:1983xr} and trace anomalies~\cite{Bastianelli:1992ct, Bastianelli:1991be}, and later it was introduced by Bern and Kosower~\cite{Bern:1990cu}, and Strassler~\cite{Strassler:1992zr}, as a proper method to compute QFT effective actions and generic QFT Feynman diagrams---see~\cite{Schubert:2001he} for a comprehensive review of the early stages of the method.  Since then, several applications and new implementations of the worldline formalism have been considered. In the realm of perturbative QFT some examples are: the computation of multiloop effective actions~\cite{Schmidt:1994aq}, Bern-Kosower rules for dressed propagators~\cite{Daikouji:1995dz, Ahmadiniaz:2015xoa}, the worldline formalism in curved spacetime~\cite{Bastianelli:2002fv, Bastianelli:2002qw, Bastianelli:2005vk, Bastianelli:2004zp},  higher-spin field theory approaches~\cite{Bastianelli:2007pv, Bastianelli:2008nm, Corradini:2010ia, Bastianelli:2012bn},  the spinning particle approach to Yang Mills theories~\cite{Dai:2008bh, Bastianelli:2013pta}, as well as applications to noncommutative QFT~\cite{Kiem:2001dk, Bonezzi:2012vr}, to the Standard Model and Grand Unified theories~\cite{Mansfield:2014vea, Edwards:2014bfa}, and to QFT on manifolds with boundary~\cite{Bastianelli:2006hq, Bastianelli:2008vh}. 

The extension of the worldline formalism to the computation of effective actions and Feynman diagrams for QFT in curved space time required to tackle some technical issues which, during several years, had resulted in numerous controversial statements and errors. The main issue boils down to the fact that, when the metric is non-flat, the associated particle models are characterized by non-linear sigma models which, in the perturbative path integral approach about the flat space metric, give rise to an infinite set of vertices with double-derivative interactions. By a simple power counting analysis, these interactions can be shown to lead to ultraviolet divergences, at the one- and two-loop level, which need to be suitably regularized.~\footnote{One-dimensional non-linear sigma models are super-renormalizable theories and diagrams with more than two loops are finite.} By now all the ambiguities have been dispelled, various regularization schemes have been devised and tested, and the method has been consistently used in several computations  (see ref.~\cite{Bastianelli:2006rx, Bastianelli:2011cc} for a detailed description of the method and for a complete list of references)---in the present work we adopt Dimensional Regularization (DR) to take care of the ambiguous diagrams. However, due to the aforementioned vertices, the computational difficulty becomes fastly harder as the order in the perturbative expansion increases, and finding simplified methods to handle the perturbative expansion in curved space would certainly be helpful, which is one of the objectives of the present manuscript.

In this paper we study bosonic particle path integrals in curved space through the computation of trace anomalies for scalar fields in various dimensions. As reviewed in sect.~\ref{se:TA}, trace anomalies are linked to particle path integrals in curved space with periodic boundary conditions, {\it i.e.} path integrals over coordinate trajectories that have the topology of a circle. In the perturbative approach, as we will see, this leaves the possibility of choosing different boundary conditions for the particle propagator, which correspond to different ways of factoring out a zero mode of the free kinetic term. Here we use the so-called ``string-inspired'' (SI) Feynman rules which correspond to the zero mode identified as  the center of mass of the paths. Along with this, we will make use of Riemann Normal Coordinates in the expansion, and specialize ourselves to maximally symmetric (MS) spaces. In curved spaces, as it is reviewed in sect.~\ref{sec:BRST}, the use of special coordinates comes with a prize: the need of a map, which we refer to as the ``geodesic map'', that for boundary conditions different than Dirichlet's gives non-trivial contributions to the perturbative expansion; as explained by Friedan~\cite{Friedan:1980jm}, this is due to the fact that a certain linear shift symmetry becomes non-linear when expressed in RNC's. In sect.~\ref{sec:Q} we thus compute the geodesic map, in closed form ({\it i.e.} to any order in the curvature), for MS spaces. Finally, in sect.~\ref{sec:TA} we obtain the type-A trace anomalies for conformal scalar field theories in MS space-times of dimension six and smaller, and test that our results reproduce known results. In sect.~\ref{sec:concl} we draw some conclusions and discuss possible extensions and applications of the model.  A technical appendix is added at the end, which includes the list of worldline integrals needed in the computation, along with a detailed example where the rules of DR are reviewed.  
 
\section{Trace anomalies in the worldline representation}
\label{se:TA}
Trace anomalies are linked to the (lack of) Weyl invariance of the effective action of a classically Weyl-invariant quantum field theory. In particular, as originally shown by Fujikawa~\cite{Fujikawa:1980vr}, in the field theory path integral approach, the trace anomaly can be seen to arise as a non-trivial Jacobian of the measure under Weyl transformations. As a paradigmatic example,  let us consider a Weyl-invariant scalar field theory $\phi(x)$ in a $D$-dimensional curved space-time, whose Wick-rotated Euclidean action reads
\begin{equation}
S[\phi,g_{\mu\nu};\xi]=\frac{1}{2}\int d^Dx\;\sqrt{g} \big( g^{\mu\nu}\partial_{\mu}\phi\partial_{\nu}\phi-\xi R\phi^2 \big),
\label{eq:ac-scal}
\end{equation}
where $\xi\defeq\frac{D-2}{4(D-1)}$ sets the non-minimal conformal coupling.~\footnote{Our conventions for the Riemann and Ricci tensors are  $[\nabla_{\mu},\nabla_{\nu}]V^{\rho}=R_{\mu\nu}{}^{\rho}{}_{\sigma}V^{\sigma}$ and $R_{\mu\nu}=R_{\mu\rho}{}^{\rho}{}_{\nu}$.} The latter is invariant under the (infinitesimal) Weyl-transformation
\begin{equation}
\delta_\sigma g_{\mu\nu}(x)=\sigma(x)g_{\mu\nu}(x),\quad \delta_\sigma \phi(x)=\frac{1}{2}\left(1-\frac{D}{2}\right)\sigma(x)\phi(x).
\end{equation}

 The one-loop gravitational effective action $\Gamma[g_{\mu\nu}]$ associated to the classical action~\eqref{eq:ac-scal} can be obtained from the functional integral
\begin{equation}
e^{-\Gamma[g_{\mu\nu}]}=\int\mathcal D\phi\;e^{-S[\phi,g_{\mu\nu};\xi]},
\label{eq:Gamma}
\end{equation}
and, under the Weyl rescaling, it  gives
\begin{align}
-\delta_\sigma \Gamma = \int d^Dx  \sigma(x)g^{\mu\nu}\frac{\delta \Gamma}{\delta g^{\mu\nu}} =\int d^Dx \sqrt{g} \frac12 \sigma(x)
\Big\langle T^\mu{}_\mu(x)\Big\rangle.
\end{align} 
 Now, in order to compute the Weyl rescaling of the r.h.s. of expression ~\eqref{eq:Gamma}  it is best to rewrite the fields in terms of the  so-called Fujikawa variables $\phi(x)\rightarrow\tilde{\phi}(x)\defeq g^{\frac{1}{4}}(x)\phi(x)$,
in order to have a dimensional-independent field transformation $\delta\tilde{\phi}=\frac{1}{2}\sigma\tilde{\phi}$. Such transformation provides a Jacobian which differs from unity by the trace of an infinite dimensional operator (the scalar field action instead is Weyl invariant by assumption)
\begin{equation}
\det\frac{\partial\tilde{\phi}'(x)}{\partial\tilde{\phi}(y)}-1=\tr\frac{\partial\delta\tilde{\phi}(x)}{\partial\tilde{\phi}(y)}=\tr\Big[ \frac{1}{2}\sigma(x)\delta^D(x-y) \Big].
\label{2}
\end{equation}
 The trace must thus be regulated
\begin{align}
\tr\Big[ \frac{1}{2}\sigma(x)\delta^D(x-y) \Big] := \lim_{\beta \to 0} \tr \Big[ \frac{1}{2}\sigma e^{-\beta {\cal R}} \Big],
\end{align}
 with the consistent regulator ${\cal R}$  being the kinetic operator of $\tilde\phi$, which reads (see~\cite{Bastianelli:2006rx} for details)
\begin{align}
{\cal R}=-\frac{1}{2}{g}^{-\frac{1}{4}}\partial_{\mu}{g}^{\frac{1}{2}}{g}^{\mu\nu}\partial_{\nu}{g}^{-\frac{1}{4}}-\frac{1}{2}\xi R.
\label{eq:R}
\end{align}
Thus,  using the identification $p_\mu=-i\partial_\mu$, the differential operator~\eqref{eq:R} can be interpreted as the quantum hamiltonian of a non-relativistic particle in curved space~\footnote{Such identification is guaranteed by the fact that, in terms of the rescaled fields, the Hilbert space inner product is given by $\langle \psi| \varphi\rangle = \int d^Dx \tilde\psi^* (x) \tilde \varphi(x)$.} 
\begin{equation}
{H}=\frac{1}{2}{g}^{-\frac{1}{4}}{p}_{\mu}{g}^{\frac{1}{2}}{g}^{\mu\nu}{p}_{\nu}{g}^{-\frac{1}{4}}-\frac{1}{2}\xi R,
\label{sec:H}
\end{equation}
and the regulated trace can be written as a particle path integral transition amplitude with periodic boundary conditions. Hence, putting all together we have,
\begin{align}
\int d^Dx \sqrt{g} \, \sigma(x) \Big\langle T^\mu{}_\mu(x)\Big\rangle = \lim_{\beta\to 0}\int_{\mathrm{PBC}}\mathcal Dx\;\sigma(x)\;e^{-S[x]},
\end{align}
where 
\begin{equation}
S[x]=\frac{1}{\beta}\int_{-1}^0dt\Big\{ \frac{1}{2}g_{\mu\nu}(x)\dot{x}^{\mu}\dot{x}^{\nu}+\beta^2\big[ V(x)+V_{\text{DR}}(x) \big] \Big\},\quad V\defeq-\frac{1}{2}\xi R
\label{eq:particle-action}
\end{equation}
is the particle action associated to the hamiltonian~\eqref{sec:H}, and $V_{\text{DR}}(x)$ is the counterterm that arises from the regularization that we choose to be Dimensional Regularization (DR), whose application to finite-time one-dimensional non-linear sigma models was proposed in~\cite{Bastianelli:2000nm}, after earlier applications to the infinite-time counterparts had been obtained~\cite{Kleinert:1999aq}. In the expressions above the limit $\beta\to 0$ is meant to convey the information that only the $\beta$-independent terms are retained---in fact, it can be shown that terms that diverge in that limit can be removed by adding local counterterms to the field theory action. Finally, by setting $\sigma$ to a constant, we recognize that
\begin{align}
\int d^Dx \sqrt{g} \,  \Big\langle T^\mu{}_\mu(x)\Big\rangle = \lim_{\beta\to 0} Z(\beta),
\label{eq:int-TA}
\end{align}
with
\begin{align}
Z(\beta):= \int_{\mathrm{PBC}}\mathcal Dx\; e^{-S[x]},
\label{eq:PF}
\end{align}
{\it i.e.} the integrated trace anomaly coincides with the $\beta$-independent part of the particle partition function.

\section{BRST-methods for the particle path integral}
\label{sec:BRST}
In the short-$\beta$ perturbative expansion of the partition function~\eqref{eq:PF}, needed to compute~\eqref{eq:int-TA}, it is convenient to expand the background metric $g_{\mu\nu}(x)$ that characterizes the action~\eqref{eq:particle-action}, around a fixed point and treat the potential and the terms with metric derivatives as perturbations. Thus, the leading term becomes a free kinetic term whose corresponding operator has a zero mode on the circle, which is related to the constant translational symmetry. This zero mode must be factored out, and an efficient way of doing  that, which we review here, was described in~\cite{Bastianelli:2003bg}. Firstly, it amounts  to decompose the generic periodic path $x^{\mu}(\tau)$ into a constant zero mode $x_0^{\mu}$ and a quantum fluctuation $y^{\mu}(\tau)$
\begin{equation}
x^{\mu}(\tau)=x_0^{\mu}+y^{\mu}(\tau).
\end{equation}
This splitting obviously introduces a constant shift symmetry
\begin{equation}
\begin{split}
\delta x_0^{\mu}&=\epsilon^{\mu}\\\delta y^{\mu}(\tau)&=-\epsilon^{\mu},
\end{split}
\label{12}
\end{equation}
which---treating both fields $x_0^{\mu}$ and $y^{\mu}(\tau)$ as dynamical variables of the path integral---behaves as a gauge symmetry. Hence, the path integral needs to be gauge-fixed in order not to overcount equivalent field configurations. This can be achieved using BRST methods: the shift symmetry~\eqref{12} is thus turned into a BRST symmetry
\begin{equation}
\begin{split}
\delta x_0^{\mu}&=\eta^{\mu}\mathit{\mathit{\Lambda}},\qquad\delta y^{\mu}(\tau)=-\eta^{\mu}\mathit{\Lambda},\\\delta \eta^{\mu}&=0,\qquad\qquad\;\:\delta \bar{\eta}_{\mu}=i\pi_{\mu}\mathit{\Lambda},\\\delta\pi_{\mu}&=0,
\end{split}
\end{equation}
where $\mathit{\Lambda}$ is an anticommuting parameter and $\eta^{\mu}$, $\bar{\eta}_{\mu}$ and $\pi_{\mu}$ are constant fields, the first two anticommuting and the third commuting.  The gauge can thus be fixed by introducing a  ``gauge fixing fermion''
\begin{equation}
\mathit{\Psi}[\rho]=\bar{\eta}_{\mu}\int_{-1}^0d\tau\;\rho(\tau)y^{\mu}(\tau),
\end{equation}
which is parameterized by a distribution $\rho(\tau)$,  normalized to $\int_{-1}^0 d\tau\rho(\tau) =1$. The gauge-fixed action reads
\begin{equation}
\begin{split}
S_{gf}[x_0,y,\eta,\bar{\eta},\pi]&\defeq S[x_0,y]+\frac{\delta}{\delta\mathit{\Lambda}}\mathit{\Psi}\\&=S[x_0,y]+i\pi_{\mu}\int_{-1}^0d\tau\;\rho(\tau)y^{\mu}(\tau)-\bar{\eta}_{\mu}\eta^{\mu}
\end{split}
\end{equation}
and all the fields (that appear as arguments of $S_{gf}$) are path-integrated. In particular, the integral over the anticommuting constant fields is equal to unity, whereas the integral over the auxiliary commuting field $\pi_\mu$ imposes the constraint
\begin{equation}
\int_{-1}^0d\tau\;\rho(\tau)y^{\mu}(\tau)=0\quad \Longrightarrow\quad \int_{-1}^0d\tau\;\rho(\tau)x^{\mu}(\tau)=x_0^\mu,
\end{equation}
which allows to invert the free kinetic operator of the fluctuations $y^\mu$, to find the particle propagator. Obviously, different gauge functions $\rho$'s give rise to different propagators, but the $\rho$-independence of the partition function is guaranteed by BRST symmetry, whereas the partition function density may, in general, be $\rho$-dependent. We can thus write the partition function as an integral over the zero mode
\begin{align}
Z(\beta) =\int d^Dx_0 \sqrt{g(x_0)}\, {\cal Z}^{(\rho)}(x_0,\beta),
\label{eq:part-fn-d}
\end{align}   
where ${\cal Z}^{(\rho)}(x_0,\beta)$ is the partition function density,  whose $\beta$-independent part yields the trace anomaly.  Moreover, the dependence on $\rho$ of the partition function density must arrange in the form of covariant total derivatives, which are indeed trivial anomalies, that can be removed by adding local counterterms to the field theory action. 

In the present calculation we use the string-inspired (SI) Feynman rules, which correspond to the choice  $\rho(\tau)=1$, where  the zero mode plays the role of the ``center of mass'' of the loop and the quantum fluctuations are periodic and have vanishing center of mass~\cite{Schubert:2001he}. Another popular choice in this type of computations is $\rho(\tau)=\delta(\tau)$ which leads to Dirichlet boundary conditions (DBC) for the fluctuations and the zero mode is the initial(=final) point of the loop. The advantage of the SI choice is that, unlike with DBC, the worldline propagator is translationally invariant. However, as we shall shortly see, in a special coordinate system, SI requires the inclusion of further vertices than DBC.  We will make use of (geodesic) Riemann Normal Coordinates (RNC) $\xi^\mu$  centered around the zero mode $x_0^\mu$, {\it i.e.}
 \begin{equation}
y^{\mu}=\xi^{\mu}-\sum_{n=2}^{\infty}\frac{1}{n!}\Gamma^{\mu}{}_{(\nu_1\nu_2;\nu_3\ldots\nu_n)}(x_0)\xi^{\nu_1}\ldots\xi^{\nu_n},
\label{eq:RNC-c}
\end{equation}
 where $\Gamma^{\mu}{}_{(\nu_1\nu_2;\nu_3\ldots\nu_n)}(x_0)$ is the symmetrized derivative of Christoffel's symbol evaluated at $x_0$, covariantized on the lower indices, which leads to RNC expansion of the metric
 \begin{equation}
\begin{split}
g_{\mu\nu}(x_0,\xi)=\;&g_{\mu\nu}(x_0)+\frac{1}{3}R_{\mu\rho\sigma\nu}(x_0)\xi^{\rho}\xi^{\sigma}+\frac{1}{6}R_{\mu\rho\sigma\nu;\alpha_1}(x_0)\xi^{\rho}\xi^{\sigma}\xi^{\alpha_1}+\\&+\Big( \frac{1}{20}R_{\mu\rho\sigma\nu;\alpha_1\alpha_2}(x_0)+\frac{2}{45}R_{\mu\rho\sigma}{}^{\beta_1}R_{\beta_1\alpha_1\alpha_2\nu}(x_0) \Big)\xi^{\rho}\xi^{\sigma}\xi^{\alpha_1}\xi^{\alpha_2}+o(\xi^5).
\end{split}
\end{equation}
Thus, the coordinate transformation~\eqref{eq:RNC-c}, induces the following non linear BRST transformation on the RNC coordinates
 \begin{align}
&\delta\xi^{\mu}(\tau)=-Q^{\mu}{}_{\nu}(x_0,\xi(\tau))\eta^{\nu}\mathit{\Lambda}, \\&
\quad Q^{\mu}{}_{\nu}(x_0,0) =\delta^\mu_\nu.\label{eq:bcQ}
\end{align}
We refer to $Q^\mu{}_\nu$ as the ``geodesic map'' and a geometric interpretation thereof is given in the following section, along with a derivation in closed form, for the case of maximally symmetric backgrounds. However, let us check here how the particle action changes if we use $\xi$ as dynamical variables.  In this case it is convenient---in strict analogy to what discussed above for a generic coordinate set---to consider the gauge-fixing fermion
\begin{equation}
\mathit{\Psi}=\bar{\eta}_{\mu}\int_{-1}^0d\tau\;\rho(\tau)\xi^{\mu}(\tau),
\end{equation}
which then yields the gauge-fixed action
\begin{equation}
\begin{split}
S_{gf}[x_0,\xi,\eta,\bar{\eta},\pi]&=S[x_0,\xi]+\frac{\delta}{\delta\mathit{\Lambda}}\mathit{\Psi}\\&=S[x_0,\xi]+i\pi_{\mu}\int_{-1}^0d\tau\;\rho(\tau)\xi^{\mu}(\tau)-\bar{\eta}_{\mu}\int_{-1}^0d\tau\;\rho(\tau)Q^{\mu}{}_{\nu}(x_0,\xi(\tau))\eta^{\nu}.
\end{split}
\label{eq:gf-action}
\end{equation}
Note that, as a consequence of condition~\eqref{eq:bcQ}, the last term of the previous expression is $\xi$-independent if $\rho(\tau) = \delta(\tau)$  and thus, for DBC, it does not introduce addition interactions. On the other hand, for SI it is  $\xi$-dependent and does introduce a new interacting piece of action, which in the perturbative approach leads to an infinite set of vertices, which must be taken into account in order to correctly compute the short-$\beta$ expansion, and ultimately the trace anomalies. Specifically, we thus get   
\begin{equation}
\begin{split}
S_{gf}[x_0,\xi,\eta,\bar{\eta},\pi]&=S[x_0,\xi]+\frac{\delta}{\delta\mathit{\Lambda}}\mathit{\Psi}\\&=S[x_0,\xi]+i\pi_{\mu}\int_{-1}^0d\tau\;\xi^{\mu}(\tau)-\bar{\eta}_{\mu}\int_{-1}^0d\tau\;Q^{\mu}{}_{\nu}(x_0,\xi(\tau))\eta^{\nu}\;.
\end{split}
\label{eq:gf-action-SI}
\end{equation}
and the Einstein-invariant and BRST-invariant path integral measure reads
\begin{equation}
\begin{split}
\mathcal Dx=dx_0\;d\eta\;d\bar{\eta}\;d\pi\prod_{-1\leq\tau<0}\sqrt{g(x_0,\xi(\tau))}\;d\xi(\tau).
\end{split}
\label{eq:PI-measure}
\end{equation}
The $\sqrt{g}$ factor of \eqref{eq:PI-measure} can now be conveniently exponentiated by introducing a set of ghost fields, $a(\tau)$ (bosonic) and $b(\tau),c(\tau)$ (fermionic), with their own dynamics~\cite{Bastianelli:1992ct},
\begin{equation}
\begin{split}
\sqrt{g(x_0,\xi(\tau))}&=\int\mathcal Da\mathcal Db \mathcal Dc\;e^{-S_{gh}}\\S_{gh}[\xi,a,b,c]&=\frac{1}{\beta}\int_{-1}^0d\tau \Big[ \frac{1}{2}g_{\mu\nu}(x_0,\xi)(a^{\mu}a^{\nu}+b^{\mu}c^{\nu}) \Big],
\end{split}
\end{equation}
so that the final quantum action is given by
\begin{equation}
S_q[x_0,\xi,\eta,\bar{\eta},\pi,a,b,c]\defeq S_{gf}[x_0,\xi,\eta,\bar{\eta},\pi]+S_{gh}[x_0,\xi,a,b,c]\;.
\end{equation}
Putting all together, the full transition amplitude reads
\begin{align}
Z(\beta)&=\int dx_0\;\sqrt{g(x_0)}\mathcal Z^{(SI)}(x_0,\beta)\nonumber \\
 &=\int dx_0d\bar{\eta}d\eta d\pi\int\mathcal D\xi\mathcal Da\mathcal Db\mathcal Dc\;e^{-S_q}.
\end{align}
In order to compute the perturbative expansion of the latter, we consider the expansion of the metric and of the geodesic map about the point $x_0^\mu$, i.e. $\xi^\mu=0$. The terms quadratic   in the various fields yield the propagators
\begin{equation}
\begin{split}
\big\langle \xi^{\mu}(\tau)\xi^{\nu}(\sigma) \big\rangle=\;&-\beta g^{\mu\nu}(x_0)\mathcal B(\tau,\sigma)\\\big\langle a^{\mu}(\tau)a^{\nu}(\sigma) \big\rangle=\;&\beta g^{\mu\nu}(x_0)\Delta_{gh}(\tau,\sigma)\\\big\langle b^{\mu}(\tau)c^{\nu}(\sigma) \big\rangle=\;&-2\beta g^{\mu\nu}(x_0)\Delta_{gh}(\tau,\sigma)\\\big\langle \bar{\eta}^{\mu}\eta_{\nu} \big\rangle=\;&\delta^{\mu}{}_{\nu},
\end{split}
\label{eq:props}
\end{equation}
with
\begin{equation}
\begin{split}
\mathcal B(\tau,\sigma)=\;&\frac{1}{2}\lvert \tau-\sigma \rvert-\frac{1}{2}(\tau-\sigma)^2-\frac{1}{12}\\\Delta_{gh}(\tau,\sigma)=\;&\delta(\tau-\sigma).
\end{split}
\label{eq:greens}
\end{equation}
The interacting part of the action, $S_q^{(int)}$, can be obtained by replacing $g_{\mu\nu}(x_0,\xi)\ \to g_{\mu\nu}(x_0,\xi) - g_{\mu\nu}(x_0)$ inside the kinetic part of~\eqref{eq:particle-action}, and by replacing $Q^\mu{}_\nu(x_0,\xi)\ \to Q^\mu{}_\nu(x_0,\xi) -\delta^\mu_\nu$ inside the BRST ghost action (for notational simplicity, we will use reparametrization invariance in $x_0$  to set  $g_{\mu\nu}(x_0)=\delta_{\mu\nu}$).  For the partition function we thus get
\begin{equation}
 Z(\beta)=\int d^Dx_0\frac{\sqrt{g(x_0)}}{(2\pi\beta)^{\frac{D}{2}}}\, \Big\langle e^{-S_q^{(int)}} \Big\rangle_{(SI)},
\label{eq:Zbeta}
\end{equation}
where the suffix $SI$ is meant to remind that we are using String-Inspired Feynman rules. Hence, comparing with~\eqref{eq:part-fn-d}, we get
\begin{align}
\mathcal Z^{(SI)}(x_0,\beta) = \frac{1}{(2\pi\beta)^{\frac{D}{2}}}\, \Big\langle e^{-S_q^{(int)}}\Big\rangle_{(SI)}
\label{eq:part-fn-SI}
\end{align}
and
\begin{align}
\Big\langle T^\mu{}_\mu (x_0)\Big\rangle =\lim_{\beta \to 0} \mathcal Z^{(SI)}(x_0,\beta)
\label{eq:local-tr-an}
\end{align}	
gives the local ({\it i.e.} unintegrated) trace anomaly at point $x_0$. 

Before proceeding further with the perturbative computation, we need to evaluate the expansion of the geodesic map $Q^{\mu}{}_{\nu}$  to the necessary order: this was discussed by Friedan in~\cite{Friedan:1980jm}. However, instead of considering a generically curved space, here we content ourselves with spaces of maximal symmetry, where 
\begin{equation}
R_{\mu\nu\rho\sigma}=b(g_{\mu\rho}g_{\nu\sigma}-g_{\mu\sigma}g_{\nu\rho}),
\label{eq:Riemann-MS}
\end{equation}
and $b:=\frac{R}{D(1-D)}$, is negative on spheres. In this case we find that the above non-linear map can be obtained in closed form. This is the subject of the following section.

\section{The geodesic map in maximally symmetric spaces}
\label{sec:Q}

As we have anticipated, in order to use RNC coordinates as quantum fluctuations in our path integral, we need to take into account that the BRST symmetry induced by the linear shift of the zero mode $x_0$, acts non linearly on the RNC's, namely  
\begin{equation}
\delta\xi^{\mu}(\tau)=-Q^{\mu}{}_{\nu}(x_0,\xi(\tau))\eta^{\nu}\mathit{\Lambda}.
\end{equation}
This stems from the fact that, by definition, $x^\mu_0$ is the origin of the RNC coordinates which are vectors on the tangent space $T_{x_0}$: they are tangent vectors, in $x^\mu_0$ to the geodesics that link $x^\mu_0$ to generic points $x^\mu$ of the manifold. Therefore, a shift of $x_0^\mu$ implies a shift of tangent space, and in turn this means that $\xi{'^\mu} =\xi^\mu +\delta \xi^\mu$ is a vector on the shifted tangent space.  Thus, if the manifold is not flat, the transformation of the RNC coordinates is a non-linear expression of the old RNC coordinates $\xi^\mu$. On the other hand, if the manifold is flat the different tangent spaces coincide and  $Q^{\mu}{}_{\nu}(x_0,\xi(\tau)) =\delta^\mu_\nu$. Moreover, if $x^\mu\equiv x^\mu_0$, {\it i.e.} $\xi^\mu =0$, then $\delta \xi^\mu =\delta y^\mu =-\delta x_0^\mu$, and $Q^\mu{}_\nu(x_0,0) =\delta^\mu_\nu$.

   Friedan~\cite{Friedan:1980jm}  proposed a method, which we briefly review below, to systematically compute the map $Q^{\mu}{}_{\nu}(x_0,\xi)$ in an arbitrary geometry as a power series in $\xi$.   Let us denote by 
\begin{equation}
\mathcal{Q}\defeq Q^{\mu}{}_{\nu}(x_0,\xi)
\label{92}
\end{equation}
the matrix which represents the geodesic map. It was found it convenient to re-write the latter in terms of another matrix $\mathcal V$, as
\begin{equation}
\mathcal Q=1+\partial \log \mathcal V.
\label{26}
\end{equation}
Above, the derivative operator is defined by
\begin{equation}
\partial\defeq\xi^{\mu}\Big(\frac{\partial}{\partial {\xi_\mu}}-\tilde\nabla_{\mu}\Big),
\end{equation}
where $\tilde\nabla_{\mu}$ is a covariant derivative that acts on tensor-valued functions of $\xi$ (for $\xi$-independent functions it reduces to the standard covariant derivative) and satisfies the property 
\begin{equation}
\tilde \nabla_{\mu}\xi^{\nu}=0.
\label{24}
\end{equation}
By formally expanding $\mathcal V$ as a power series in $\xi$
\begin{equation}
\mathcal V=\sum_{n=0}^{\infty}\frac{1}{(n+1)!}\mathcal V^{(n)},
\label{20}
\end{equation}
where
\begin{equation}
\mathcal {V}^{(0)}=\mathbbm 1,\quad\mathcal {V}^{(1)}=0,\quad\mathcal V^{(n)}\propto\xi^n,
\end{equation}
one obtains that, in a generic torsion-free space, the matrices $\mathcal V^{(n)}$ satisfy the recursion relation
\begin{equation}
\mathcal V^{(n)}=2\nabla\mathcal V^{(n-1)}-\nabla^2\mathcal V^{(n-2)}+\mathcal V^{(n-2)}\mathcal R,
\end{equation}
with
\begin{equation}
\mathcal R\defeq R^{\mu}{}_{\rho\sigma\nu}(x_0)\xi^{\rho}\xi^{\sigma}.
\label{93}
\end{equation}
The previous recursion relation uniquely fixes $\mathcal V$ order by order in $\xi$. However, by increasing the order, the calculation becomes rapidly harder and, for a generic manifold, a closed form for the matrix is not known. On the other hand, for MS spaces  we obviously have that
\begin{equation}
\tilde \nabla_{\alpha}\mathcal R=0,
\label{22}
\end{equation}
which immediately implies
\begin{align}
&\mathcal V^{(2n+1)} =0,\\
&\mathcal V^{(2n)} =\mathcal R^n 
\end{align}
and we thus get
\begin{equation}
\mathcal V=\sum^{\infty}_{n=0}\frac{1}{(2n+1)!}\mathcal R^n  =\frac{\sinh\sqrt{\mathcal R}}{\sqrt{\mathcal R}},
\label{21}
\end{equation}
with
\begin{equation}
\mathcal R^0\defeq\mathbbm 1,\quad\mathcal R^n\defeq R^{\mu}{}_{\alpha_1\beta_1\rho_1}R^{\rho_1}{}_{\alpha_2\beta_2\rho_2}\cdots R^{\rho_{n-1}}{}_{\alpha_n\beta_n\nu}(x_0)\;\xi^{\alpha_1}\xi^{\beta_1}\cdots\xi^{\alpha_n}\xi^{\beta_n}.
\label{91}
\end{equation}
Moreover, note that for MS spaces the operator $\partial$ defined above just acts as a number operator, {\it i.e.} $\partial \mathcal R = 2 \mathcal R$. Therefore, the geodesic map simply reads
\begin{align}
\mathcal Q^{(MS)}=\sqrt{\mathcal R}\coth\sqrt{\mathcal R},
\end{align} 
which can be easily expanded to the desired order. Before doing that, let us first rearrange it in a more convenient form. Note in fact that, using~\eqref{eq:Riemann-MS} and~\eqref{93}, we get
\begin{align}
\mathcal R = b (\delta^\mu_\sigma \delta_{\rho\nu} -\delta^\mu_\nu \delta_{\rho\sigma}) \xi^\rho \xi^\sigma =:-b\xi^2 \mathcal P 
\end{align}
in terms of the projector $\mathcal P = \delta^\mu_\nu -\frac{\xi^\mu\xi_\nu}{\xi^2}$, which satisfies the condition $\partial \mathcal P=0$. We thus get
\begin{align}
&\mathcal R^n =(-b\xi^2)^n \mathcal P ~,
\end{align}
and finally
\begin{align}
& \mathcal V^{(MS)} = \mathbbm 1 +\mathcal P \Biggl( \frac{\sinh\sqrt{-b\xi^2}}{\sqrt{-b\xi^2}}-1\Biggr),\\
& \mathcal Q^{(MS)} = \mathbbm 1 +\mathcal P \Bigl( \sqrt{-b\xi^2} \coth\sqrt{-b\xi^2}-1\Bigr).
\end{align}
Hence, one can easily expand the previous expression in power series of $b$. In components the expansion reads, 
\begin{equation}
\begin{split}
{Q^{(MS)}}^{\mu}{}_{\nu}=\delta^{\mu}{}_{\nu}+\left[\frac{b}{3}+\frac{b^2}{45}\xi_{\rho}\xi^{\rho}+\frac{2}{945}b^3\left( \xi_{\rho}\xi^{\rho} \right)^2+\ldots\right]
\left( \xi^{\mu}\xi_{\nu}-\delta^{\mu}{}_{\nu}\xi_{\rho}\xi^{\rho} \right).
\end{split}
\label{eq:Q-exp}
\end{equation}
Here we only keep the terms that will be needed  in the following section to perform our trace anomaly tests. 

\section{Computation of the trace anomaly}
\label{sec:TA}
In order to compute the local trace anomaly of a conformally coupled scalar field theory, we need to obtain the perturbative expansion of the correlator of eq.~\eqref{eq:part-fn-SI}, which involves the interacting quantum  action, whose derivation was explained in sect.~\ref{sec:BRST}, namely  
\begin{equation}
\begin{split}
S_q^{(int)}=&\;\frac{1}{\beta}\int_{-1}^0dt\Big\{ \frac{1}{2}\big[ g_{\mu\nu}(x_0,\xi)-\delta_{\mu\nu} \big]\big[ \dot{\xi}^{\mu}\dot{\xi}^{\nu}+a^{\mu}a^{\nu}+b^{\mu}c^{\nu} \big]+\beta^2\big[ V(x_0,\xi)+V_{DR}(x_0,\xi) \big] \Big\} +\\&\;- \bar{\eta}_{\mu}\int_{-1}^0d\tau\;\big[Q^{\mu}{}_{\nu}(x_0,\xi(\tau))-\delta^{\mu}{}_{\nu}\big]\eta^{\nu}.
\end{split}
\label{30}
\end{equation}
Notice that, in the MS geometry, the potential term $-\beta(V+V_{DR})$  is a constant and can thus be factored out from the correlator, {\it i.e.}
\begin{equation}
\mathcal Z^{(SI)}(x_0,\beta)=\frac{e^{-\beta(1-4\xi)\frac{R}{8}}}{(2\pi\beta)^{\frac{D}{2}}}\Big\langle e^{-\tilde{S}_q^{(int)}} \Big\rangle_{(SI)},
\label{43}
\end{equation}
where the new interacting quantum action is given by
\begin{equation}
\tilde{S}_q^{(int)}\defeq S_q^{(int)}-\beta(1-4\xi)\frac{R}{8}.
\label{31}
\end{equation}
In the present work we content ourselves we the computation of trace anomalies in dimension six or smaller, for which the necessary RNC expansion of the metric (in MS spaces) is already known, and can be found for instance in ref.~\cite{Bastianelli:2001tb}
\begin{equation}
g_{\mu\nu}(\xi)=\delta_{\mu\nu}+2\left( \xi_{\mu}\xi_{\nu}-\delta_{\mu\nu}\xi_{\rho}\xi^{\rho} \right)\left[ \frac{b}{6}-\frac{16}{6!}b^2\left( \xi_{\rho}\xi^{\rho} \right)^2 + \frac{8}{7!}b^3 \left( \xi_{\rho}\xi^{\rho} \right)^4+\ldots \right],
\label{eq:metric-RNC}
\end{equation}
whereas the expansion of the geodesic map is the one given above in eq.~\eqref{eq:Q-exp}.
Hence, 
\begin{align}
\tilde{S}_q^{(int)}=&\;\frac{1}{\beta}\int_{-1}^0\mathrm{d}\tau\;\Bigg[ \frac{b}{6}-\frac{16}{6!}b^2 \xi_{\rho}(\tau)\xi^{\rho}(\tau)  + \frac{8}{7!}b^3 \left[ \xi_{\rho}(\tau)\xi^{\rho}(\tau) \right]^2+\ldots \Bigg]\times\nonumber\\&\times\Bigg[ \xi_{\mu}(\tau)\xi_{\nu}(\tau)-\delta_{\mu\nu}\xi_{\rho}(\tau)\xi^{\rho}(\tau) \Bigg]\Bigg[ \dot{\xi}^{\mu}(\tau)\dot{\xi}^{\nu}(\tau)+a^{\mu}(\tau)a^{\nu}(\tau)+\nonumber\\&+b^{\mu}(\tau)c^{\nu}(\tau) \Bigg]-\bar{\eta}_{\mu}\int_{-1}^0\mathrm{d}\tau\;\Bigg[\frac{b}{3}+\frac{b^2}{45}\xi_{\rho}(\tau)\xi^{\rho}(\tau)+\nonumber\\&+\frac{2}{945}b^3\left[ \xi_{\rho}(\tau)\xi^{\rho}(\tau) \right]^2+\ldots\Bigg]\Bigg[ \xi^{\mu}(\tau)\xi_{\nu}(\tau)-\delta^{\mu}{}_{\nu}\xi_{\rho}(\tau)\xi^{\rho}(\tau) \Bigg] \eta^{\nu}.
\end{align}
Using $\beta$ as the perturbative parameter, the above action can be split up as
\begin{equation}
\tilde{S}_q^{(int)}=\underbrace{S'_2}_{\beta}+\underbrace{S_4}_{\beta}+\underbrace{S'_4}_{\beta^2}+\underbrace{S_6}_{\beta^2}+\underbrace{S'_6}_{\beta^3}+\underbrace{S_8}_{\beta^3}+\ldots,
\end{equation}
where, for each term, its perturbative weight is indicated. In particular, such terms are
{\allowdisplaybreaks
\begin{align}
S_4=&\;\frac{b}{6\beta}\int_{-1}^0\mathrm{d}\tau\;\Big[ \xi_{\mu}(\tau)\xi_{\nu}(\tau)-\delta_{\mu\nu}\xi_{\rho}(\tau)\xi^{\rho}(\tau) \Big]\Big[ \dot{\xi}^{\mu}(\tau)\dot{\xi}^{\nu}(\tau)+a^{\mu}(\tau)a^{\nu}(\tau)+\nonumber\\&\;+b^{\mu}(\tau)c^{\nu}(\tau) \Big],\\  
S_6
=&\;\frac{-16b^2}{6!\beta}\int_{-1}^0\mathrm{d}\tau\;\xi_{\sigma}(\tau)\xi^{\sigma}(\tau)\Big[ \xi_{\mu}(\tau)\xi_{\nu}(\tau)-\delta_{\mu\nu}\xi_{\rho}(\tau)\xi^{\rho}(\tau) \Big]\times\nonumber\\ 
&\;\times\left[ \dot{\xi}^{\mu}(\tau)\dot{\xi}^{\nu}(\tau)+a^{\mu}(\tau)a^{\nu}(\tau)+b^{\mu}(\tau)c^{\nu}(\tau) \right],
 \\S_8=&\;\frac{8b^3}{7!\beta}\int_{-1}^0\mathrm{d}\tau\;\xi_{\sigma}(\tau)\xi^{\sigma}(\tau)\xi_{\alpha}(\tau)\xi^{\alpha}(\tau)\Big[ \xi_{\mu}(\tau)\xi_{\nu}(\tau)-\delta_{\mu\nu}\xi_{\rho}(\tau)\xi^{\rho}(\tau) \Big]\times\nonumber
 \\ &\;\times\Big[ \dot{\xi}^{\mu}(\tau)\dot{\xi}^{\nu}(\tau)+a^{\mu}(\tau)a^{\nu}(\tau)+b^{\mu}(\tau)c^{\nu}(\tau) \Big],
 \\S'_2=&\;-\bar{\eta}_{\mu}\frac{b}{3}\int_{-1}^0\mathrm{d}\tau\;\Big[ \xi^{\mu}(\tau)\xi_{\nu}(\tau)-\delta^{\mu}{}_{\nu}\xi_{\rho}(\tau)\xi^{\rho}(\tau) \Big]\eta^{\nu},
  \\S'_4=&\;-\bar{\eta}_{\mu}\frac{b^2}{45}\int_{-1}^0\mathrm{d}\tau\;\xi_{\sigma}(\tau)\xi^{\sigma}(\tau)\Big[ \xi^{\mu}(\tau)\xi_{\nu}(\tau)-\delta^{\mu}{}_{\nu}\xi_{\rho}(\tau)\xi^{\rho}(\tau) \Big]\eta^{\nu},
  \\S'_6=&\;-\bar{\eta}_{\mu}\frac{2b^3}{945}\int_{-1}^0\mathrm{d}\tau\;\xi_{\sigma}(\tau)\xi^{\sigma}(\tau)\xi_{\alpha}(\tau)\xi^{\alpha}(\tau)\Big[ \xi^{\mu}(\tau)\xi_{\nu}(\tau)-\delta^{\mu}{}_{\nu}\xi_{\rho}(\tau)\xi^{\rho}(\tau) \Big]\eta^{\nu}
\end{align}
}
and they can be used to reduce the contraction in \eqref{43} to (to avoid cluttering we omit the suffix $SI$)
{\allowdisplaybreaks
\begin{align}
\left \langle e^{-\tilde{S}^{(int)}_q} \right\rangle=&\exp\Bigg( -\underbrace{\left\langle S_4 \right\rangle}_{\beta}-\underbrace{\left\langle S'_2 \right\rangle}_{\beta}-\underbrace{\left\langle S_6 \right\rangle}_{\beta^2}-\underbrace{\left\langle S'_4 \right\rangle}_{\beta^2}+\frac{1}{2}\underbrace{\left\langle {S_4}^2 \right\rangle_C}_{\beta^2}+\frac{1}{2}\underbrace{\left\langle {S'_2}^2 \right\rangle_C}_{\beta^2}+\nonumber\\&+\underbrace{\left\langle {S'}_2S_4 \right\rangle_C}_{\beta^2}-\underbrace{\left\langle S_8 \right\rangle}_{\beta^3}-\underbrace{\left\langle S'_6 \right\rangle}_{\beta^3}+\underbrace{\left\langle S_4S_6 \right\rangle_C}_{\beta^3}-\frac{1}{3!}\underbrace{\left\langle {S_4}^3 \right\rangle_C}_{\beta^3}-\frac{1}{3!}\underbrace{\left\langle {S'_2}^3 \right\rangle_C}_{\beta^3}+\nonumber\\&+\underbrace{\left\langle S'_2S_6 \right\rangle_C}_{\beta^3}+\underbrace{\left\langle S'_2S'_4 \right\rangle_C}_{\beta^3}+\underbrace{\left\langle S'_4S_4 \right\rangle_C}_{\beta^3}-\frac{1}{2}\underbrace{\left\langle {S'_2}^2S_4 \right\rangle_C}_{\beta^3}-\frac{1}{2}\underbrace{\left\langle S'_2{S_4}^2 \right\rangle_C}_{\beta^3}+\ldots \Bigg).
\label{44}
\end{align}
}\hspace{-1mm}
In the following, we report the results for the various contractions of \eqref{44}, expressed both in terms of their string-inspired worldline integrals (we indicate them with $\mathcal M$) and then explicitly computed---we already write them in terms of the curvature scalar $R$. The $\mathcal M$ integrals are reported in Appendix~\ref{app:B}:
{\allowdisplaybreaks
\begin{align}
\left\langle S_4 \right\rangle &=\frac{\beta R}{6}  \mathcal M_1  
=-\frac{1}{72}\beta R,\\ 
\left\langle S'_2 \right\rangle &=\frac{\beta R}{3}\mathcal M_2
= -\frac{1}{36}\beta R,\\
\left\langle S_6\right\rangle &= \frac{16\beta^2R^2}{6!D\left( 1-D \right)}(D+2)\mathcal M_3 
=-\frac{1}{6480}\frac{(D+2)}{D(D-1)}\beta^2R^2,
\\\left\langle S'_4 \right\rangle & =-\frac{\beta^2R^2}{45D\left( 1-D \right)}(D+2)\mathcal M_4
=\frac{1}{6480}\frac{(D+2)}{D(D-1)}\beta^2R^2,
\\
\left\langle {S'_2}^2 \right\rangle_C&=\frac{\beta^2R^2}{3^2D \left(D -1\right)}\left[ -(D-1)(\mathcal M_2)^2 +(2D-5)\mathcal M_5\right]
=-\frac{1}{2160}\frac{1}{D-1}\beta^2R^2,
\\
\left\langle {S_4}^2 \right\rangle_C &=\frac{\beta^2R^2}{18D \left( D -1\right)}\Big[ (D-1)(2\mathcal M_6+\mathcal M_7+\mathcal M_8)+3(\mathcal M_9-2\mathcal M_{10}+\mathcal M_{11})\Big]\nonumber\\&
=-\frac{1}{6480}\frac{7D-46}{D(D-1)}\beta^2R^2,
\\
\left \langle S'_2S_4 \right\rangle_C&=\frac{\beta^2R^2}{9D} (\mathcal M_{12}+\mathcal M_{13}) =-\frac{1}{1620}\frac{1}{D}\beta^2R^2,\\
\left\langle S_8 \right\rangle&=\frac{8\beta^3R^3}{7!D^2(1-D)^2}(D+2)(D+4)\mathcal M_{14}
=-\frac{8}{7!\cdot1728}\frac{(D+2)(D+4)}{D^2(1-D)^2}\beta^3R^3,\\
\left\langle S'_6 \right\rangle &=\frac{2}{945}\frac{\beta^3R^3}{D^2(1-D)^2}(D+2)(D+4)\mathcal M_{15}
=-\frac{2}{945\cdot1728}\frac{(D+2)(D+4)}{D^2(1-D)^2}\beta^3R^3,\\
\left\langle S_4S_6 \right\rangle_C&=\frac{16\beta^3R^3}{6\cdot6!D^2(1-D)^2}2(D+2)\Big[(D+1)(-2\mathcal M_{16}-\mathcal M_{17}-2\mathcal M_{19}-\mathcal M_{21})+\nonumber\\&+5(-\mathcal M_{18}+2\mathcal M_{20}-\mathcal M_{22})\Big]
=-\frac{16}{6\cdot6!\cdot2160}\frac{(9D-74)(D+2)}{D^2(1-D)^2}\beta^3R^3,\\
\left\langle {S_4}^3 \right\rangle_C &=-\frac{\beta^3 R^3}{6^3D^2(1-D)^2}\Big[-24(D-1)^2(\mathcal M_{23}+\mathcal M_{24}-2\mathcal M_{25}+\mathcal M_{26}+\mathcal M_{28}+\mathcal M_{29}+\nonumber\\&+\frac{1}{3}\mathcal M_{30}+\mathcal M_{32}-\mathcal M_{35}+\frac{1}{3}\mathcal M_{39}-\mathcal M_{44})-72(D-1)(\mathcal M_{27}+\mathcal M_{31}+\mathcal M_{33}-\mathcal M_{36}+\nonumber\\&-2\mathcal M_{37}+\mathcal M_{40}-\mathcal M_{41}+2\mathcal M_{43}-\mathcal M_{45}-2\mathcal M_{46}+2\mathcal M_{47}-\mathcal M_{48})-24(2D-5)(\mathcal M_{34}+\nonumber\\&-\mathcal M_{38}-\mathcal M_{49}-2\mathcal M_{50}+2\mathcal M_{51}+2\mathcal M_{52}+\frac{1}{3}\mathcal M_{55})+8(D-16)(\mathcal M_{42}+3\mathcal M_{54})+\nonumber\\&+24(D+11)(\mathcal M_{53}+\frac{1}{3}\mathcal M_{56})\Big]
=-\frac{1}{6^3\cdot7560}\frac{289D^2-2464D-4068}{D^2(1-D)^2}\beta^3R^3,\\
\left\langle {S'_2}^3 \right\rangle_C &=-\frac{\beta^3R^3}{3^3D^2(1-D)^2}\Big[ -2(D-1)^2\mathcal M_{57}+6(D-1)(2D-5)\mathcal M_{58}+\nonumber\\&-2(D-2)(4D-19)\mathcal M_{59} \Big]
=-\frac{1}{3^3\cdot30240}\frac{D^2+23D+6}{D^2(1-D)^2}\beta^3R^3,\\
\left\langle S'_2S_6 \right\rangle_C &=\frac{16\beta^3R^3}{3\cdot6! D^2(1-D)}2(D+2)(2\mathcal M_{60}+\mathcal M_{61})
=\frac{16}{3\cdot6!\cdot1440}\frac{D+2}{D^2(1-D)}\beta^3R^3,\\
\left\langle S'_2S'_4 \right\rangle_C &=\frac{\beta^3R^3}{3\cdot45D^2(1-D)^2}(D+2)\Big[ (4D-9)\mathcal M_{62}+(D-1)\mathcal M_{63} \Big]=\nonumber\\&
=\frac{1}{3\cdot45\cdot8640}\frac{(D+2)(D+4)}{D^2(1-D)^2}\beta^3R^3,\\
\left\langle S'_4S_4 \right\rangle_C &=-\frac{\beta^3R^3}{6\cdot45 D^2(1-D)^2}4\left( D+2 \right)\left( \mathcal M_{64}+\mathcal M_{65} \right)
=-\frac{112}{6\cdot45\cdot60480}\frac{D+2}{D^2(1-D)}\beta^3R^3,\\
\left\langle {S'_2}^2S_4 \right\rangle_C &=\frac{\beta^3R^3}{6\cdot3^2 D^2(1-D)^3}\Big[-4(D-1)^2(2D-5)(\mathcal M_{66}+\mathcal M_{68})-2(2D^3-10D^2+17D+\nonumber\\&-7)\mathcal M_{67}+4(D-1)(2D^2-6D+7)\mathcal M_{69}-2(2D^3-6D^2+9D-7)\mathcal M_{70}+\nonumber\\&+4(D-1)^3(\mathcal M_{71}+\mathcal M_{72}) \Big]
=-\frac{4}{54\cdot60480}\frac{4D^2+21D-46}{D^2(1-D)^2}\beta^3R^3,\\
\left\langle S'_2{S_4}^2 \right\rangle_C &=-\frac{\beta^3R^3}{3\cdot6^2D^2(1-D)}\Big[ 8(D-1)(\mathcal M_{73}+\mathcal M_{74}-2\mathcal M_{75}-2\mathcal M_{76}+\mathcal M_{77}+2\mathcal M_{81}+\nonumber\\&+\mathcal M_{83}+\mathcal M_{84}+\mathcal M_{85})+24(\mathcal M_{78}-\mathcal M_{79}-2\mathcal M_{80}+2\mathcal M_{82}+\mathcal M_{86}+\mathcal M_{87})\Big] =\nonumber\\&
=\frac{1}{3\cdot6^2\cdot7560}\frac{37D+158}{D^2(1-D)}\beta^3R^3.
\end{align}
}\hspace{-1mm}
In the above calculations, all terms containing equal time propagators with one derivative have been excluded, as they vanish---see Appendix~\ref{app:B}. This fact contributes significantly to simply the expansion in  Wick's contractions.

Now, putting all together, the local trace anomaly can be extracted from
\begin{equation}
\begin{split}
&\langle T^{\mu}{}_{\mu}(x_0) \rangle=\lim_{\beta\to0}\mathcal Z(\beta)=\lim_{\beta\to0}\frac{1}{(2\pi\beta)^{\frac{D}{2}}}\exp\Bigg[ \frac{\beta}{4!}(12\xi-2)R+\\&\quad-\frac{\beta^2}{6!}\frac{(D-3)}{D(D-1)}R^2+\frac{\beta^3}{8!}\frac{16(D+2)(D-3)}{9D^2(D-1)^2}R^3+\ldots \Bigg].
\end{split}
\label{46}
\end{equation}
At fixed dimension $D$, the $\beta$-limit selects the $\beta$-independent part in the expansion of the exponent in~\eqref{46} after the simplification with the Feynman measure ${1}/{(2\pi\beta)^{D/2}}$, whereas $\beta$-divergent terms are ignored as they may be removed by a QFT renormalization procedure. Recalling that $\xi=\frac{D-2}{4(D-1)}$, the result of our trace anomaly  reads
\begin{equation}
\begin{split}
D=2&\quad\Longrightarrow\quad\left\langle T^{\mu}{}_{\mu}(x_0) \right\rangle=-\frac{R}{24\,\pi}\\
D=4&\quad\Longrightarrow\quad\left\langle T^{\mu}{}_{\mu}(x_0) \right\rangle=-\frac{R^2}{48\cdot6!\,\pi^2}\\
D=6&\quad\Longrightarrow\quad\left\langle T^{\mu}{}_{\mu}(x_0) \right\rangle=-\frac{R^3}{60\cdot9!\,\pi^3},
\end{split}
\end{equation}
which is in perfect agreement with the results obtained using the standard DBC procedure~\cite{Bastianelli:2001tb}. Note that trivial anomalies are absent in MS spaces, as they would appear as covariant derivatives of curvature combinations.

\section{Conclusions}
\label{sec:concl}
We have discussed the application of the  string-inspired method within the worldline formalism in curved space which, on the circle,  allows to compute one-loop effective actions and associated scattering amplitudes, and anomalies. The implementation of SI Feynman rules corresponds to a convenient way of factoring out a zero mode present on the circle. A BRST technique, studied in~\cite{Bastianelli:2003bg}, has been used for that purpose, along with RNC coordinates.

 The main advantage of using the SI Feynman rules, in place of those associated to different ways of factoring out the zero mode (such as DBC), resides in the simplicity of the worldline propagator which results translationally invariant, unlike the DBC propagator. Therefore, all the diagrams that involve equal time propagators with one derivative, are vanishing.   The price to pay for such advantage is the introduction of further vertices in the theory, which arise from a non-linear ``geodesic map''.  Above we have computed such map in closed form, for MS spaces, and successfully tested the associated non-linear sigma model via the computation of type-A trace anomalies of conformally-coupled scalar fields in dimension not larger than six. 
 
 The  string-inspired formalism in curved space can be exploited in a wider class of calculations,  and can be considered as a powerful tool to reduce the complexity of standard Feynman diagrams computations, in quantum field theory and in string theory~\cite{Green:2008bf, Basu:2018eep}. One important example is the systematic computation of graviton scattering amplitudes and gravitational effective actions. It is an outstanding problem, and the development of new methods, both analytic and numeric, may be of considerable help. To such extent, an interesting scenario, that was conjectured years ago in~\cite{Guven:1987en} and recently improved in~\cite{Bastianelli:2017wsy, Bastianelli:2017xhi, Bastianelli:2018twb}, consists in the possibility of mapping the particle non-linear sigma model to a linear sigma model where the gravitational properties are described in terms of an effective potential, with a substantial gain of effectiveness in the perturbative computation. So far, such mapping has been studied only with DBC Feynman rules and seems to be  guaranteed only for MS spaces. However, it would be interesting to investigate the possibility of using the SI method there since, because of flatness, the geodesic map should not add complications.  Another relevant extension involves supersymmetric sigma models, which are linked to the worldline approach for Dirac particles in curved space~\cite{Bastianelli:2002qw}, where it is certainly possible to consider SI Feynman rules.

\appendix
\section{Worldline integrals}
\label{app:B}

The worldline integrals, that enter in the perturbative calculation described in the main text, involve the coordinate Green's function $\mathcal B$ and the ghost Green's function  $\Delta_{\text{gh}}$, which read
\begin{align}
& \mathcal B(\tau,\sigma) =\mathcal B(\sigma,\tau)=\frac{1}{2}\lvert \tau-\sigma \rvert-\frac{1}{2}(\tau-\sigma)^2-\frac{1}{12}
\label{eq:particle-prop}
\\
& \Delta_{\text{gh}}(\tau,\sigma) = \delta(\tau-\sigma),
\end{align}
and derivatives of the former, which at the unregulated level read
\begin{align}
\leftidx{^{\bullet}}{\mathcal{B}(\tau,\sigma)}{}&=\frac{1}{2}\text{sgn}(\tau-\sigma)-\tau+\sigma=-\mathcal{B}^{\bullet}(\tau,\sigma)\\ \leftidx{^{\bullet \bullet}}{\mathcal{B}(\tau,\sigma)}{}&=\delta(\tau-\sigma)-1=\mathcal{B}^{\bullet \bullet}(\tau,\sigma).
\end{align}
Due to the translational invariance of the string inspired propagator~\eqref{eq:particle-prop}, the derivative with respect to the second variable (right bullet) is the opposite of the derivative with respect to the first variable (left bullet). However, for future convenience, in the formulas below we prefer to keep explicit the distinction. 
Propagators satisfy the properties
\begin{equation}
\begin{split}
\int_{-1}^0\mathrm{d}\tau\;\mathcal{B}(\tau,\sigma)=\left( \leftidx{^{\bullet}}{\mathcal{B}(\tau,\sigma)}{} \right)\big\rvert_{\sigma=\tau}&=0,\\
 \Big[\leftidx{^{\bullet\hspace{-0.1cm}}}{\mathcal{B}^\bullet(\tau,\sigma)}{} +  \Delta_{\text{gh}}(\tau,\sigma)\Big]_{\sigma=\tau}&=1,
\end{split}
\label{45}
\end{equation}
which will be largely exploited in the actual computation. The last property of \eqref{45} shows an example of divergence cancellation: a $\delta(0)$ term gets canceled in the sum. This is the simplest example of how the ghost fields contribute to cancel worldline divergences.

In the following, we report the list of the SI worldline integrals which have been used for the calculation of the trace anomalies. They are computed using dimensional regularization (DR), when needed. For completeness, at the end of the section we provide an example of how DR works in this worldline context.
To simplify the notation, we define $\int \defeq \int_{-1}^0$, $\mathcal B\vert_{\tau}\defeq \mathcal B(\tau,\tau)$ and $\mathcal B \defeq \mathcal B(\tau_1,\tau_2)$ (or $\mathcal B_{12}\defeq\mathcal B(\tau_1,\tau_2)$ for triple integrals). 
{\allowdisplaybreaks
\begin{align*}
\mathcal M_1&=\int\mathrm d\tau\;\mathcal B\rvert_{\tau}\left( \leftidx{^{\bullet\hspace{-0.1cm}}}{\mathcal B}{^{\bullet}}+\Delta_{\text{gh}} \right)\rvert_{\tau}=-\frac{1}{12}\\
\mathcal M_2&=\int\mathrm d\tau\;\mathcal B\rvert_{\tau}=-\frac{1}{12}
\\
\mathcal M_3&=\int\mathrm d\tau\;\mathcal B\rvert_{\tau}{}^2\left( \leftidx{^{\bullet\hspace{-0.1cm}}}{\mathcal B}{^{\bullet}}+\Delta_{\text{gh}} \right)\rvert_{\tau}=\frac{1}{144}
\\
\mathcal M_4 &=\int\mathrm d\tau\;\mathcal B\rvert_{\tau}{}^2=\frac{1}{144}\\
\mathcal M_5&=\int\mathrm d\tau_1\int\mathrm d\tau_2\;\mathcal B^2=\frac{1}{720}\\
\mathcal M_6&=\int\mathrm d\tau_1\int\mathrm d\tau_2\;\mathcal B\rvert_1\;\leftidx{^{\bullet\hspace{-0.1cm}}}{\mathcal B}{}{}^2\left( \leftidx{^{\bullet\hspace{-0.1cm}}}{\mathcal B}{^{\bullet}}+\Delta_{\text{gh}} \right)\rvert_2=-\frac{1}{144}\\
\mathcal M_7&=\int\mathrm d\tau_1\int\mathrm d\tau_2\;\mathcal B\rvert_1\mathcal B\rvert_2\left( \leftidx{^{\bullet\hspace{-0.1cm}}}{\mathcal B}{^{\bullet}}{}^2-\Delta_{\text{gh}}{}^2 \right)=-\frac{1}{144}\\
\mathcal M_8&=\int\mathrm d\tau_1\int\mathrm d\tau_2\;\mathcal B^2\left( \leftidx{^{\bullet\hspace{-0.1cm}}}{\mathcal B}{^{\bullet}}+\Delta_{\text{gh}} \right)\rvert_1\left( \leftidx{^{\bullet\hspace{-0.1cm}}}{\mathcal B}{^{\bullet}}+\Delta_{\text{gh}} \right)\rvert_2=\frac{1}{720}\\
\mathcal M_9&=\int\mathrm d\tau_1\int\mathrm d\tau_2\;\mathcal B^2\left( \leftidx{^{\bullet\hspace{-0.1cm}}}{\mathcal B}{^{\bullet}}{}^2-\Delta_{\text{gh}}{}^2 \right)=\frac{1}{120}\\
\mathcal M_{10}&=\int\mathrm d\tau_1\int\mathrm d\tau_2\;\mathcal B\;\leftidx{}{\mathcal B}{^{\bullet}}\;\leftidx{^{\bullet\hspace{-0.1cm}}}{\mathcal B}{}\;\leftidx{^{\bullet\hspace{-0.1cm}}}{\mathcal B}{^{\bullet}}=-\frac{11}{1440}\\
\mathcal M_{11}&=\int\mathrm d\tau_1\int\mathrm d\tau_2\;\leftidx{}{\mathcal B}{^{\bullet}}{}^2\;\leftidx{^{\bullet\hspace{-0.1cm}}}{\mathcal B}{}{}^2=\frac{1}{80}\\
\mathcal M_{12}&=\int\mathrm d\tau_1\int\mathrm d\tau_2\;\mathcal B^2\left( \leftidx{^{\bullet\hspace{-0.1cm}}}{\mathcal B}{^{\bullet}}+\Delta_{\text{gh}} \right)\rvert_2=\frac{1}{720}\\
\mathcal M_{13}&=\int\mathrm d\tau_1\int\mathrm d\tau_2\;\leftidx{}{\mathcal B}{^{\bullet}}{}^2\mathcal B\rvert_2=-\frac{1}{144}\\
\mathcal M_{14}&=\int\mathrm d\tau\;\mathcal B\rvert_{\tau}{}^3\left( \leftidx{^{\bullet\hspace{-0.1cm}}}{\mathcal B}{^{\bullet}}+\Delta_{\text{gh}} \right)\rvert_{\tau}=-\frac{1}{1728}\\
\mathcal M_{15}&=\int\mathrm d\tau\;\mathcal B\rvert_{\tau}{}^3=-\frac{1}{1728}
\\
\mathcal M_{16}&=\int\mathrm d\tau_1\int\mathrm d\tau_2\;\mathcal B\rvert_1\mathcal B\rvert_2\leftidx{^{\bullet\hspace{-0.1cm}}}{\mathcal B}{}{}^2\left( \leftidx{^{\bullet\hspace{-0.1cm}}}{\mathcal B}{^{\bullet}}+\Delta_{\text{gh}} \right)\rvert_2=\frac{1}{1728}\\
\mathcal M_{17}&=\int\mathrm d\tau_1\int\mathrm d\tau_2\;\mathcal B\rvert_1\mathcal B\rvert_2{}^2\left( \leftidx{^{\bullet\hspace{-0.1cm}}}{\mathcal B}{^{\bullet}}{}^2-\Delta_{\text{gh}}{}^2 \right)=\frac{1}{1728}\\
\mathcal M_{18}&=\int\mathrm d\tau_1\int\mathrm d\tau_2\;\mathcal B^2\mathcal B\rvert_2\left( \leftidx{^{\bullet\hspace{-0.1cm}}}{\mathcal B}{^{\bullet}}{}^2-\Delta_{\text{gh}}{}^2 \right)=-\frac{1}{1440}\\
\mathcal M_{19}&=\int\mathrm d\tau_1\int\mathrm d\tau_2\;\mathcal B^2\mathcal B\rvert_2\left( \leftidx{^{\bullet\hspace{-0.1cm}}}{\mathcal B}{^{\bullet}}+\Delta_{\text{gh}} \right)\rvert_1\left( \leftidx{^{\bullet\hspace{-0.1cm}}}{\mathcal B}{^{\bullet}}+\Delta_{\text{gh}} \right)\rvert_2=-\frac{1}{8640}\\
\mathcal M_{20}&=\int\mathrm d\tau_1\int\mathrm d\tau_2\;\mathcal B\rvert_2\;\mathcal B\;\leftidx{}{\mathcal B}{^{\bullet}}\;\leftidx{^{\bullet\hspace{-0.1cm}}}{\mathcal B}{}\;\leftidx{^{\bullet\hspace{-0.1cm}}}{\mathcal B}{^{\bullet}}=\frac{11}{17280}\\
\mathcal M_{21}&=\int\mathrm d\tau_1\int\mathrm d\tau_2\;\leftidx{}{\mathcal B}{^{\bullet}}{}^2\mathcal B\rvert_2{}^2\left( \leftidx{^{\bullet\hspace{-0.1cm}}}{\mathcal B}{^{\bullet}}+\Delta_{\text{gh}} \right)\rvert_1=\frac{1}{1728}\\
\mathcal M_{22}&=\int\mathrm d\tau_1\int\mathrm d\tau_2\;\leftidx{}{\mathcal B}{^{\bullet}}{}^2\;\leftidx{^{\bullet\hspace{-0.1cm}}}{\mathcal B}{}{}^2\mathcal B\rvert_2=-\frac{1}{960}\\
\mathcal M_{23}&=\int\mathrm d\tau_1\int\mathrm d\tau_2\int\mathrm d\tau_3\;\mathcal B\rvert_1\leftidx{^{\bullet\hspace{-0.1cm}}}{\mathcal B_{12}}{}{}^2\;\leftidx{^{\bullet\hspace{-0.1cm}}}{\mathcal B_{23}}{}{}^2\left( \leftidx{^{\bullet\hspace{-0.1cm}}}{\mathcal B}{^{\bullet}}+\Delta_{\text{gh}} \right)\rvert_3=-\frac{1}{1728}
\\
\mathcal M_{24}&=\int\mathrm d\tau_1\int\mathrm d\tau_2\int\mathrm d\tau_3\;\mathcal B\rvert_1{}^2\;\leftidx{^{\bullet\hspace{-0.1cm}}}{\mathcal B_{12}}{}{}^2\left( \leftidx{^{\bullet\hspace{-0.1cm}}}{\mathcal B_{23}}{^{\bullet}}{}^2-\Delta_{\text{gh},23}{}^2 \right)=-\frac{1}{1728}\\
\mathcal M_{25}&=\int\mathrm d\tau_1\int\mathrm d\tau_2\int\mathrm d\tau_3\;\mathcal B\rvert_1\;\leftidx{^{\bullet\hspace{-0.1cm}}}{\mathcal B_{12}}{}\;\mathcal B_{23}\;\leftidx{^{\bullet\hspace{-0.1cm}}}{\mathcal B_{23}}{}\;\leftidx{^{\bullet\hspace{-0.1cm}}}{\mathcal B_{12}}{^{\bullet}}=0\\
\mathcal M_{26}&=\int\mathrm d\tau_1\int\mathrm d\tau_2\int\mathrm d\tau_3\;\mathcal B\rvert_1\;\leftidx{^{\bullet\hspace{-0.1cm}}}{\mathcal B_{12}}{}\;\leftidx{^{\bullet\hspace{-0.1cm}}}{\mathcal B_{13}}{}\;\mathcal B_{23}\left( \leftidx{^{\bullet\hspace{-0.1cm}}}{\mathcal B}{^{\bullet}}+\Delta_{\text{gh}} \right)\rvert_2\left( \leftidx{^{\bullet\hspace{-0.1cm}}}{\mathcal B}{^{\bullet}}+\Delta_{\text{gh}} \right)\rvert_3=\frac{1}{8640}\\
\mathcal M_{27}&=\int\mathrm d\tau_1\int\mathrm d\tau_2\int\mathrm d\tau_3\;\mathcal B\rvert_1\;\leftidx{^{\bullet\hspace{-0.1cm}}}{\mathcal B_{12}}{}\;\leftidx{^{\bullet\hspace{-0.1cm}}}{\mathcal B_{13}}{}\;\mathcal B_{23}\left( \leftidx{^{\bullet\hspace{-0.1cm}}}{\mathcal B_{23}}{^{\bullet}}{}^2-\Delta_{\text{gh},23}{}^2 \right)=\frac{1}{1440}\\
\mathcal M_{28}&=\int\mathrm d\tau_1\int\mathrm d\tau_2\int\mathrm d\tau_3\;\mathcal B\rvert_1{}^2\;\leftidx{^{\bullet\hspace{-0.1cm}}}{\mathcal B_{12}}{}\;\leftidx{}{\mathcal B_{23}}{^{\bullet}}\;\leftidx{^{\bullet\hspace{-0.1cm}}}{\mathcal B_{13}}{^{\bullet}}\left( \leftidx{^{\bullet\hspace{-0.1cm}}}{\mathcal B}{^{\bullet}}+\Delta_{\text{gh}} \right)\rvert_2=-\frac{1}{1728}\\
\mathcal M_{29}&=\int\mathrm d\tau_1\int\mathrm d\tau_2\int\mathrm d\tau_3\;\mathcal B\rvert_1\;\mathcal B_{23}{}^2\;\left( \leftidx{^{\bullet\hspace{-0.1cm}}}{\mathcal B_{12}}{^{\bullet}}{}^2-\Delta_{\text{gh},12}{}^2 \right)\left( \leftidx{^{\bullet\hspace{-0.1cm}}}{\mathcal B}{^{\bullet}}+\Delta_{\text{gh}} \right)\rvert_3=\frac{1}{8640}\\
\mathcal M_{30}&=\int\mathrm d\tau_1\int\mathrm d\tau_2\int\mathrm d\tau_3\;\mathcal B\rvert_1{}^3\;\left( \leftidx{^{\bullet\hspace{-0.1cm}}}{\mathcal B_{12}}{^{\bullet}}\;\leftidx{^{\bullet\hspace{-0.1cm}}}{\mathcal B_{23}}{^{\bullet}}\;\leftidx{^{\bullet\hspace{-0.1cm}}}{\mathcal B_{13}}{^{\bullet}}+\Delta_{\text{gh},12}\;\Delta_{\text{gh},23}\;\Delta_{\text{gh},13} \right)=-\frac{1}{1728}\\
\mathcal M_{31}&=\int\mathrm d\tau_1\int\mathrm d\tau_2\int\mathrm d\tau_3\;\mathcal B\rvert_1\;\mathcal B_{23}{}^2\left( \leftidx{^{\bullet\hspace{-0.1cm}}}{\mathcal B_{12}}{^{\bullet}}\;\leftidx{^{\bullet\hspace{-0.1cm}}}{\mathcal B_{23}}{^{\bullet}}\;\leftidx{^{\bullet\hspace{-0.1cm}}}{\mathcal B_{13}}{^{\bullet}}+\Delta_{\text{gh},12}\;\Delta_{\text{gh},23}\;\Delta_{\text{gh},13} \right)=\frac{1}{1440}\\
\mathcal M_{32}&=\int\mathrm d\tau_1\int\mathrm d\tau_2\int\mathrm d\tau_3\;\mathcal B_{12}{}^2\;\leftidx{^{\bullet\hspace{-0.1cm}}}{\mathcal B_{23}}{}{}^2\left( \leftidx{^{\bullet\hspace{-0.1cm}}}{\mathcal B}{^{\bullet}}+\Delta_{\text{gh}} \right)\rvert_1\left( \leftidx{^{\bullet\hspace{-0.1cm}}}{\mathcal B}{^{\bullet}}+\Delta_{\text{gh}} \right)\rvert_3=\frac{1}{8640}\\
\mathcal M_{33}&=\int\mathrm d\tau_1\int\mathrm d\tau_2\int\mathrm d\tau_3\;\mathcal B_{12}{}^2\;\leftidx{^{\bullet\hspace{-0.1cm}}}{\mathcal B_{13}}{}\;\leftidx{^{\bullet\hspace{-0.1cm}}}{\mathcal B_{23}}{}\;\leftidx{^{\bullet\hspace{-0.1cm}}}{\mathcal B_{12}}{^{\bullet}}\left( \leftidx{^{\bullet\hspace{-0.1cm}}}{\mathcal B}{^{\bullet}}+\Delta_{\text{gh}} \right)\rvert_3=-\frac{11}{20160}\\
\mathcal M_{34}&=\int\mathrm d\tau_1\int\mathrm d\tau_2\int\mathrm d\tau_3\;\mathcal B_{12}{}^2\;\leftidx{^{\bullet\hspace{-0.1cm}}}{\mathcal B_{13}}{}{}^2\left( \leftidx{^{\bullet\hspace{-0.1cm}}}{\mathcal B_{23}}{^{\bullet}}{}^2-\Delta_{\text{gh},23}{}^2 \right)=-\frac{1}{4032}
\\
\mathcal M_{35}&=\int\mathrm d\tau_1\int\mathrm d\tau_2\int\mathrm d\tau_3\;\mathcal B_{12}\;\leftidx{}{\mathcal B_{12}}{^{\bullet}}\;\mathcal B_{23}\;\leftidx{^{\bullet\hspace{-0.1cm}}}{\mathcal B_{23}}{}\left( \leftidx{^{\bullet\hspace{-0.1cm}}}{\mathcal B}{^{\bullet}}+\Delta_{\text{gh}} \right)\rvert_1\left( \leftidx{^{\bullet\hspace{-0.1cm}}}{\mathcal B}{^{\bullet}}+\Delta_{\text{gh}} \right)\rvert_3=0\\
\mathcal M_{36}&=\int\mathrm d\tau_1\int\mathrm d\tau_2\int\mathrm d\tau_3\;\mathcal B_{12}\;\leftidx{}{\mathcal B_{12}}{^{\bullet}}\;\leftidx{^{\bullet\hspace{-0.1cm}}}{\mathcal B_{12}}{}\;\leftidx{^{\bullet\hspace{-0.1cm}}}{\mathcal B_{13}}{}\;\leftidx{^{\bullet\hspace{-0.1cm}}}{\mathcal B_{23}}{}\left( \leftidx{^{\bullet\hspace{-0.1cm}}}{\mathcal B}{^{\bullet}}+\Delta_{\text{gh}} \right)\rvert_3=\frac{11}{60480}\\
\mathcal M_{37}&=\int\mathrm d\tau_1\int\mathrm d\tau_2\int\mathrm d\tau_3\;\mathcal B_{12}\;\leftidx{}{\mathcal B_{12}}{^{\bullet}}\;\leftidx{^{\bullet\hspace{-0.1cm}}}{\mathcal B_{13}}{}\;\mathcal B_{23}\;\leftidx{^{\bullet\hspace{-0.1cm}}}{\mathcal B_{12}}{^{\bullet}}\left( \leftidx{^{\bullet\hspace{-0.1cm}}}{\mathcal B}{^{\bullet}}+\Delta_{\text{gh}} \right)\rvert_3=-\frac{1}{60480}\\
\mathcal M_{38}&=\int\mathrm d\tau_1\int\mathrm d\tau_2\int\mathrm d\tau_3\;\mathcal B_{12}\;\leftidx{}{\mathcal B_{12}}{^{\bullet}}\;\mathcal B_{23}\;\leftidx{^{\bullet\hspace{-0.1cm}}}{\mathcal B_{23}}{}\left( \leftidx{^{\bullet\hspace{-0.1cm}}}{\mathcal B_{13}}{^{\bullet}}{}^2-\Delta_{\text{gh},13}{}^2 \right)=\frac{61}{120960}\\
\mathcal M_{39}&=\int\mathrm d\tau_1\int\mathrm d\tau_2\int\mathrm d\tau_3\;\mathcal B_{12}\;\mathcal B_{23}\;\mathcal B_{13}\;\left( \leftidx{^{\bullet\hspace{-0.1cm}}}{\mathcal B}{^{\bullet}}+\Delta_{\text{gh}} \right)\rvert_1\left( \leftidx{^{\bullet\hspace{-0.1cm}}}{\mathcal B}{^{\bullet}}+\Delta_{\text{gh}} \right)\rvert_2\left( \leftidx{^{\bullet\hspace{-0.1cm}}}{\mathcal B}{^{\bullet}}+\Delta_{\text{gh}} \right)\rvert_3=-\frac{1}{30240}\\
\mathcal M_{40}&=\int\mathrm d\tau_1\int\mathrm d\tau_2\int\mathrm d\tau_3\;\mathcal B_{12}\;\mathcal B_{23}\;\mathcal B_{13}\;\left( \leftidx{^{\bullet\hspace{-0.1cm}}}{\mathcal B}{^{\bullet}}+\Delta_{\text{gh}} \right)\rvert_1\left( \leftidx{^{\bullet\hspace{-0.1cm}}}{\mathcal B_{23}}{^{\bullet}}{}^2-\Delta_{\text{gh},23}{}^2 \right)=\frac{1}{40320}\\
\mathcal M_{41}&=\int\mathrm d\tau_1\int\mathrm d\tau_2\int\mathrm d\tau_3\;\mathcal B_{12}\;\mathcal B_{13}\;\leftidx{}{\mathcal B_{23}}{^{\bullet}}\;\leftidx{^{\bullet\hspace{-0.1cm}}}{\mathcal B_{23}}{}\;\leftidx{^{\bullet\hspace{-0.1cm}}}{\mathcal B_{23}}{^{\bullet}}\left( \leftidx{^{\bullet\hspace{-0.1cm}}}{\mathcal B}{^{\bullet}}+\Delta_{\text{gh}} \right)\rvert_1=\frac{13}{120960}\\
\mathcal M_{42}&=\int\mathrm d\tau_1\int\mathrm d\tau_2\int\mathrm d\tau_3\;\mathcal B_{12}\;\mathcal B_{23}\;\mathcal B_{13}\;\left( \leftidx{^{\bullet\hspace{-0.1cm}}}{\mathcal B_{12}}{^{\bullet}}\;\leftidx{^{\bullet\hspace{-0.1cm}}}{\mathcal B_{23}}{^{\bullet}}\;\leftidx{^{\bullet\hspace{-0.1cm}}}{\mathcal B_{13}}{^{\bullet}}+\Delta_{\text{gh},12}\;\Delta_{\text{gh},23}\;\Delta_{\text{gh},13} \right)=\frac{143}{120960}\\
\mathcal M_{43}&=\int\mathrm d\tau_1\int\mathrm d\tau_2\int\mathrm d\tau_3\;\mathcal B_{12}\;\leftidx{}{\mathcal B_{13}}{^{\bullet}}\;\leftidx{}{\mathcal B_{23}}{^{\bullet}}\;\leftidx{^{\bullet\hspace{-0.1cm}}}{\mathcal B_{23}}{}{}^2
\left( \leftidx{^{\bullet\hspace{-0.1cm}}}{\mathcal B}{^{\bullet}}+\Delta_{\text{gh}} \right)\rvert_1=-\frac{1}{6720}
\\
\mathcal M_{44}&=\int\mathrm d\tau_1\int\mathrm d\tau_2\int\mathrm d\tau_3\;\mathcal B\rvert_1\;\mathcal B\rvert_3\;\leftidx{^{\bullet\hspace{-0.1cm}}}{\mathcal B_{12}}{}\;\leftidx{}{\mathcal B_{23}}{^{\bullet}}\;\leftidx{^{\bullet\hspace{-0.1cm}}}{\mathcal B_{12}}{^{\bullet}}\;\leftidx{^{\bullet\hspace{-0.1cm}}}{\mathcal B_{23}}{^{\bullet}}=0\\
\mathcal M_{45}&=\int\mathrm d\tau_1\int\mathrm d\tau_2\int\mathrm d\tau_3\;\mathcal B\rvert_1\;\leftidx{^{\bullet\hspace{-0.1cm}}}{\mathcal B_{12}}{}\;\leftidx{^{\bullet\hspace{-0.1cm}}}{\mathcal B_{13}}{}\;\leftidx{}{\mathcal B_{23}}{^{\bullet}}\;\leftidx{^{\bullet\hspace{-0.1cm}}}{\mathcal B_{23}}{}\;\leftidx{^{\bullet\hspace{-0.1cm}}}{\mathcal B_{23}}{^{\bullet}}=-\frac{11}{17280}\\
\mathcal M_{46}&=\int\mathrm d\tau_1\int\mathrm d\tau_2\int\mathrm d\tau_3\;\mathcal B\rvert_1\;\leftidx{^{\bullet\hspace{-0.1cm}}}{\mathcal B_{12}}{}\;\mathcal B_{23}\;\leftidx{^{\bullet\hspace{-0.1cm}}}{\mathcal B_{23}}{}\;\leftidx{^{\bullet\hspace{-0.1cm}}}{\mathcal B_{13}}{^{\bullet}}\;\leftidx{^{\bullet\hspace{-0.1cm}}}{\mathcal B_{23}}{^{\bullet}}=-\frac{11}{17280}\\
\mathcal M_{47}&=\int\mathrm d\tau_1\int\mathrm d\tau_2\int\mathrm d\tau_3\;\mathcal B\rvert_1\;\leftidx{^{\bullet\hspace{-0.1cm}}}{\mathcal B_{12}}{}\;\leftidx{}{\mathcal B_{23}}{^{\bullet}}\;\leftidx{^{\bullet\hspace{-0.1cm}}}{\mathcal B_{23}}{}{}^2\;\leftidx{^{\bullet\hspace{-0.1cm}}}{\mathcal B_{13}}{^{\bullet}}=\frac{1}{960}\\
\mathcal M_{48}&=\int\mathrm d\tau_1\int\mathrm d\tau_2\int\mathrm d\tau_3\;\mathcal B\rvert_1\;\mathcal B_{23}\;\leftidx{}{\mathcal B_{23}}{^{\bullet}}\;\leftidx{^{\bullet\hspace{-0.1cm}}}{\mathcal B_{23}}{}\;\leftidx{^{\bullet\hspace{-0.1cm}}}{\mathcal B_{12}}{^{\bullet}}\;\leftidx{^{\bullet\hspace{-0.1cm}}}{\mathcal B_{13}}{^{\bullet}}=-\frac{11}{17280}\\
\mathcal M_{49}&=\int\mathrm d\tau_1\int\mathrm d\tau_2\int\mathrm d\tau_3\;\mathcal B_{12}{}^2\;\leftidx{^{\bullet\hspace{-0.1cm}}}{\mathcal B_{13}}{}\;\leftidx{^{\bullet\hspace{-0.1cm}}}{\mathcal B_{23}}{}\;\leftidx{^{\bullet\hspace{-0.1cm}}}{\mathcal B_{13}}{^{\bullet}}\;\leftidx{^{\bullet\hspace{-0.1cm}}}{\mathcal B_{23}}{^{\bullet}}=\frac{1}{30240}\\
\mathcal M_{50}&=\int\mathrm d\tau_1\int\mathrm d\tau_2\int\mathrm d\tau_3\;\mathcal B_{12}\;\leftidx{}{\mathcal B_{12}}{^{\bullet}}\;\leftidx{^{\bullet\hspace{-0.1cm}}}{\mathcal B_{13}}{}{}^2\;\leftidx{}{\mathcal B_{23}}{^{\bullet}}\;\leftidx{^{\bullet\hspace{-0.1cm}}}{\mathcal B_{23}}{^{\bullet}}=\frac{1}{30240}\\
\mathcal M_{51}&=\int\mathrm d\tau_1\int\mathrm d\tau_2\int\mathrm d\tau_3\;\mathcal B_{12}\;\leftidx{}{\mathcal B_{12}}{^{\bullet}}\;\leftidx{^{\bullet\hspace{-0.1cm}}}{\mathcal B_{13}}{}\;\mathcal B_{23}\;\leftidx{^{\bullet\hspace{-0.1cm}}}{\mathcal B_{13}}{^{\bullet}}\;\leftidx{^{\bullet\hspace{-0.1cm}}}{\mathcal B_{23}}{^{\bullet}}=-\frac{79}{120960}\\
\mathcal M_{52}&=\int\mathrm d\tau_1\int\mathrm d\tau_2\int\mathrm d\tau_3\;\mathcal B_{12}\;\leftidx{}{\mathcal B_{12}}{^{\bullet}}\;\leftidx{^{\bullet\hspace{-0.1cm}}}{\mathcal B_{13}}{}\;\leftidx{^{\bullet\hspace{-0.1cm}}}{\mathcal B_{23}}{}\;\leftidx{}{\mathcal B_{23}}{^{\bullet}}\;\leftidx{^{\bullet\hspace{-0.1cm}}}{\mathcal B_{13}}{^{\bullet}}=-\frac{1}{30240}\\
\mathcal M_{53}&=\int\mathrm d\tau_1\int\mathrm d\tau_2\int\mathrm d\tau_3\;\mathcal B_{12}\;\mathcal B_{13}\;\leftidx{^{\bullet\hspace{-0.1cm}}}{\mathcal B_{23}}{}\;\leftidx{}{\mathcal B_{23}}{^{\bullet}}\;\leftidx{^{\bullet\hspace{-0.1cm}}}{\mathcal B_{12}}{^{\bullet}}\;\leftidx{^{\bullet\hspace{-0.1cm}}}{\mathcal B_{13}}{^{\bullet}}=-\frac{19}{40320}
\\
\mathcal M_{54}&=\int\mathrm d\tau_1\int\mathrm d\tau_2\int\mathrm d\tau_3\;\mathcal B_{12}\;\leftidx{}{\mathcal B_{13}}{^{\bullet}}\;\leftidx{^{\bullet\hspace{-0.1cm}}}{\mathcal B_{12}}{^{\bullet}}\;\leftidx{^{\bullet\hspace{-0.1cm}}}{\mathcal B_{13}}{}\;\leftidx{}{\mathcal B_{23}}{^{\bullet}}\;\leftidx{^{\bullet\hspace{-0.1cm}}}{\mathcal B_{23}}{}=\frac{11}{12096}\\
\mathcal M_{55}&=\int\mathrm d\tau_1\int\mathrm d\tau_2\int\mathrm d\tau_3\;\leftidx{}{\mathcal B_{12}}{^{\bullet}}{}^2\;\leftidx{^{\bullet\hspace{-0.1cm}}}{\mathcal B_{13}}{}{}^2\;\leftidx{}{\mathcal B_{23}}{^{\bullet}}{}^2=\frac{17}{20160}\\
\mathcal M_{56}&=\int\mathrm d\tau_1\int\mathrm d\tau_2\int\mathrm d\tau_3\;\leftidx{}{\mathcal B_{12}}{^{\bullet}}\;\leftidx{}{\mathcal B_{13}}{^{\bullet}}\;\leftidx{^{\bullet\hspace{-0.1cm}}}{\mathcal B_{12}}{}\;\leftidx{^{\bullet\hspace{-0.1cm}}}{\mathcal B_{13}}{}\;\leftidx{}{\mathcal B_{23}}{^{\bullet}}\;\leftidx{^{\bullet\hspace{-0.1cm}}}{\mathcal B_{23}}{}=-\frac{17}{20160}\\
\mathcal M_{57}&=\int\mathrm d\tau_1\int\mathrm d\tau_2\int\mathrm d\tau_3\;\mathcal B\rvert_1\;\mathcal B\rvert_2\;\mathcal B\rvert_3=-\frac{1}{1728}\\
\mathcal M_{58}&=\int\mathrm d\tau_1\int\mathrm d\tau_2\int\mathrm d\tau_3\;\mathcal B\rvert_1\;\mathcal B_{23}{}^2=-\frac{1}{8640}\\
\mathcal M_{59}&=\int\mathrm d\tau_1\int\mathrm d\tau_2\int\mathrm d\tau_3\;\mathcal B_{12}\;\mathcal B_{23}\;\mathcal B_{13}=-\frac{1}{30240}\\
\mathcal M_{60}&=\int\mathrm d\tau_1\int\mathrm d\tau_2\;\mathcal B^2\;\mathcal B\rvert_2\left( \leftidx{^{\bullet\hspace{-0.1cm}}}{\mathcal B}{^{\bullet}}+\Delta_{\text{gh}} \right)\rvert_2=-\frac{1}{8640}\\
\mathcal M_{61}&=\int\mathrm d\tau_1\int\mathrm d\tau_2\;\leftidx{}{\mathcal B}{^{\bullet}}{}^2\;\mathcal B\rvert_2=\frac{1}{1728}\\
\mathcal M_{62}&=\int\mathrm d\tau_1\int\mathrm d\tau_2\;\mathcal B\rvert_2\;\mathcal B^2=-\frac{1}{8640}\\
\mathcal M_{63}&=\int\mathrm d\tau_1\int\mathrm d\tau_2\;\mathcal B\rvert_1\;B\rvert_2{}^2=-\frac{1}{1728}
\\
\mathcal M_{64}&=\int\mathrm d\tau_1\int\mathrm d\tau_2\;\mathcal B\rvert_1\;\mathcal B\rvert_2\;\leftidx{^{\bullet\hspace{-0.1cm}}}{\mathcal B}{}{}^2=\frac{1}{1728}\\
\mathcal M_{65}&=\int\mathrm d\tau_1\int\mathrm d\tau_2\;\mathcal B^2\;\mathcal B\rvert_2\left( \leftidx{^{\bullet\hspace{-0.1cm}}}{\mathcal B}{^{\bullet}}+\Delta_{\text{gh}} \right)\rvert_1=-\frac{1}{8640}\\
\mathcal M_{66}&=\int\mathrm d\tau_1\int\mathrm d\tau_2\int\mathrm d\tau_3\;\mathcal B\rvert_1\;\leftidx{^{\bullet\hspace{-0.1cm}}}{\mathcal B_{12}}{}\;\leftidx{^{\bullet\hspace{-0.1cm}}}{\mathcal B_{13}}{}\;\mathcal B_{23}=\frac{1}{8640}\\
\mathcal M_{67}&=\int\mathrm d\tau_1\int\mathrm d\tau_2\int\mathrm d\tau_3\;\mathcal B_{12}{}^2\;\leftidx{^{\bullet\hspace{-0.1cm}}}{\mathcal B_{13}}{}{}^2=\frac{1}{8640}\\
\mathcal M_{68}&=\int\mathrm d\tau_1\int\mathrm d\tau_2\int\mathrm d\tau_3\;\mathcal B_{12}\;\mathcal B_{23}\;\mathcal B_{13}\;\left( \leftidx{^{\bullet\hspace{-0.1cm}}}{\mathcal B}{^{\bullet}}+\Delta_{\text{gh}} \right)\rvert_1=-\frac{1}{30240}\\
\mathcal M_{69}&=\int\mathrm d\tau_1\int\mathrm d\tau_2\int\mathrm d\tau_3\;\mathcal B_{12}\;\mathcal B_{13}\;\leftidx{^{\bullet\hspace{-0.1cm}}}{\mathcal B_{12}}{}\;\leftidx{^{\bullet\hspace{-0.1cm}}}{\mathcal B_{13}}{}=0\\
\mathcal M_{70}&=\int\mathrm d\tau_1\int\mathrm d\tau_2\int\mathrm d\tau_3\;\mathcal B_{13}{}^2\;\leftidx{^{\bullet\hspace{-0.1cm}}}{\mathcal B_{12}}{}{}^2=\frac{1}{8640}\\
\mathcal M_{71}&=\int\mathrm d\tau_1\int\mathrm d\tau_2\int\mathrm d\tau_3\;\mathcal B\rvert_1\;\mathcal B\rvert_3\;\leftidx{^{\bullet\hspace{-0.1cm}}}{\mathcal B_{12}}{}{}^2=\frac{1}{1728}
\\
\mathcal M_{72}&=\int\mathrm d\tau_1\int\mathrm d\tau_2\int\mathrm d\tau_3\;\mathcal B_{12}{}^2\;\mathcal B\rvert_3\left( \leftidx{^{\bullet\hspace{-0.1cm}}}{\mathcal B}{^{\bullet}}+\Delta_{\text{gh}} \right)\rvert_1=-\frac{1}{8640}\\
\mathcal M_{73}&=\int\mathrm d\tau_1\int\mathrm d\tau_2\int\mathrm d\tau_3\;\mathcal B_{12}{}^2\;\leftidx{^{\bullet\hspace{-0.1cm}}}{\mathcal B_{23}}{}{}^2\left( \leftidx{^{\bullet\hspace{-0.1cm}}}{\mathcal B}{^{\bullet}}+\Delta_{\text{gh}} \right)\rvert_3=\frac{1}{8640}\\
\mathcal M_{74}&=\int\mathrm d\tau_1\int\mathrm d\tau_2\int\mathrm d\tau_3\;\mathcal B\rvert_3\;\mathcal B_{12}{}^2\left( \leftidx{^{\bullet\hspace{-0.1cm}}}{\mathcal B_{23}}{^{\bullet}}{}^2-\Delta_{\text{gh},23}{}^2 \right)=\frac{1}{8640}\\
\mathcal M_{75}&=\int\mathrm d\tau_1\int\mathrm d\tau_2\int\mathrm d\tau_3\;\mathcal B_{12}\;\leftidx{}{\mathcal B_{12}}{^{\bullet}}\;\mathcal B_{23}\;\leftidx{^{\bullet\hspace{-0.1cm}}}{\mathcal B_{23}}{}\left( \leftidx{^{\bullet\hspace{-0.1cm}}}{\mathcal B}{^{\bullet}}+\Delta_{\text{gh}} \right)\rvert_3=0\\
\mathcal M_{76}&=\int\mathrm d\tau_1\int\mathrm d\tau_2\int\mathrm d\tau_3\;\mathcal B_{12}\;\leftidx{}{\mathcal B_{12}}{^{\bullet}}\;\leftidx{}{\mathcal B_{23}}{^{\bullet}}\;\leftidx{^{\bullet\hspace{-0.1cm}}}{\mathcal B_{23}}{^{\bullet}}\;\mathcal B\rvert_3=0\\
\mathcal M_{77}&=\int\mathrm d\tau_1\int\mathrm d\tau_2\int\mathrm d\tau_3\;\mathcal B_{12}\;\mathcal B_{23}\;\mathcal B_{13}\left( \leftidx{^{\bullet\hspace{-0.1cm}}}{\mathcal B}{^{\bullet}}+\Delta_{\text{gh}} \right)\rvert_2\left( \leftidx{^{\bullet\hspace{-0.1cm}}}{\mathcal B}{^{\bullet}}+\Delta_{\text{gh}} \right)\rvert_3=-\frac{1}{30240}\\
\mathcal M_{78}&=\int\mathrm d\tau_1\int\mathrm d\tau_2\int\mathrm d\tau_3\;\mathcal B_{12}\;\mathcal B_{23}\;\mathcal B_{13}\left( \leftidx{^{\bullet\hspace{-0.1cm}}}{\mathcal B_{23}}{^{\bullet}}{}^2-\Delta_{\text{gh},23}{}^2 \right)=\frac{1}{40320}\\
\mathcal M_{79}&=\int\mathrm d\tau_1\int\mathrm d\tau_2\int\mathrm d\tau_3\;\mathcal B_{12}\;\mathcal B_{13}\;\leftidx{^{\bullet\hspace{-0.1cm}}}{\mathcal B_{23}}{}\;\leftidx{}{\mathcal B_{23}}{^{\bullet}}\;\leftidx{^{\bullet\hspace{-0.1cm}}}{\mathcal B_{23}}{^{\bullet}}=\frac{13}{120960}\\
\mathcal M_{80}&=\int\mathrm d\tau_1\int\mathrm d\tau_2\int\mathrm d\tau_3\;\mathcal B_{12}\;\leftidx{}{\mathcal B_{13}}{^{\bullet}}\;\mathcal B_{23}\;\leftidx{^{\bullet\hspace{-0.1cm}}}{\mathcal B_{23}}{}\;\leftidx{^{\bullet\hspace{-0.1cm}}}{\mathcal B_{23}}{^{\bullet}}=-\frac{1}{60480}\\
\mathcal M_{81}&=\int\mathrm d\tau_1\int\mathrm d\tau_2\int\mathrm d\tau_3\;\mathcal B_{12}\;\leftidx{}{\mathcal B_{13}}{^{\bullet}}\;\leftidx{}{\mathcal B_{23}}{^{\bullet}}\;\mathcal B\rvert_3\left( \leftidx{^{\bullet\hspace{-0.1cm}}}{\mathcal B}{^{\bullet}}+\Delta_{\text{gh}} \right)\rvert_2=\frac{1}{8640}\\
\mathcal M_{82}&=\int\mathrm d\tau_1\int\mathrm d\tau_2\int\mathrm d\tau_3\;\mathcal B_{12}\;\leftidx{}{\mathcal B_{13}}{^{\bullet}}\;\leftidx{}{\mathcal B_{23}}{^{\bullet}}\;\leftidx{^{\bullet\hspace{-0.1cm}}}{\mathcal B_{23}}{}{}^2=-\frac{1}{6720}
\\
\mathcal M_{83}&=\int\mathrm d\tau_1\int\mathrm d\tau_2\int\mathrm d\tau_3\;\leftidx{}{\mathcal B_{12}}{^{\bullet}}{}^2\;\mathcal B_{23}{}^2\left( \leftidx{^{\bullet\hspace{-0.1cm}}}{\mathcal B}{^{\bullet}}+\Delta_{\text{gh}} \right)\rvert_3=\frac{1}{8640}\\
\mathcal M_{84}&=\int\mathrm d\tau_1\int\mathrm d\tau_2\int\mathrm d\tau_3\;\leftidx{}{\mathcal B_{12}}{^{\bullet}}{}^2\;\leftidx{}{\mathcal B_{23}}{^{\bullet}}{}^2\;\mathcal B\rvert_3=-\frac{1}{1728}\\
\mathcal M_{85}&=\int\mathrm d\tau_1\int\mathrm d\tau_2\int\mathrm d\tau_3\;\mathcal B\rvert_2\;\mathcal B\rvert_3\;\leftidx{}{\mathcal B_{12}}{^{\bullet}}\;\leftidx{}{\mathcal B_{13}}{^{\bullet}}\;\leftidx{^{\bullet\hspace{-0.1cm}}}{\mathcal B_{23}}{^{\bullet}}=-\frac{1}{1728}\\
\mathcal M_{86}&=\int\mathrm d\tau_1\int\mathrm d\tau_2\int\mathrm d\tau_3\;\mathcal B_{23}{}^2\;\leftidx{}{\mathcal B_{12}}{^{\bullet}}\;\leftidx{}{\mathcal B_{13}}{^{\bullet}}\;\leftidx{^{\bullet\hspace{-0.1cm}}}{\mathcal B_{23}}{^{\bullet}}=-\frac{11}{20160}\\
\mathcal M_{87}&=\int\mathrm d\tau_1\int\mathrm d\tau_2\int\mathrm d\tau_3\;\leftidx{}{\mathcal B_{12}}{^{\bullet}}\;\leftidx{}{\mathcal B_{13}}{^{\bullet}}\;\mathcal B_{23}\;\leftidx{}{\mathcal B_{23}}{^{\bullet}}\;\leftidx{^{\bullet\hspace{-0.1cm}}}{\mathcal B_{23}}{}=\frac{11}{60480}.
\end{align*}
}\hspace{-1mm}
To provide an example, we report a step-by-step DR calculation of the integral $\mathcal M_{27}$.
\begin{equation*}
\begin{split}
\mathcal M_{27}=&\int\mathrm d\tau_1\int\mathrm d\tau_2\int\mathrm d\tau_3\;\mathcal B\rvert_1\;\leftidx{^{\bullet\hspace{-0.1cm}}}{\mathcal B_{12}}{}\;\leftidx{^{\bullet\hspace{-0.1cm}}}{\mathcal B_{13}}{}\;\mathcal B_{23}\left( \leftidx{^{\bullet\hspace{-0.1cm}}}{\mathcal B_{23}}{^{\bullet}}{}^2-\Delta_{\text{gh},23}{}^2 \right)\rightarrow\\&\rightarrow\int\mathrm d^{D+1}t_1\int\mathrm d^{D+1}t_2\int\mathrm d^{D+1}t_3\;\mathcal B\rvert_1\;\leftidx{_{\mu}}{\mathcal B_{12}}{}\;\leftidx{_{\mu}}{\mathcal B_{13}}{}\;\mathcal B_{23}\left( \leftidx{_{\nu}}{\mathcal B_{23}}{}{_{\rho}}{}^2-\Delta_{\text{gh},23}{}^2 \right)=
\end{split}
\end{equation*}
\begin{equation}
\begin{split}
=&\;\mathcal B\rvert_1\iiint\leftidx{_{\mu}}{\mathcal B_{12}}{}\;\leftidx{_{\mu}}{\mathcal B_{13}}{}\;\mathcal B_{23}\left( \leftidx{_{\nu}}{\mathcal B}{_{23\rho}}\;\leftidx{_{\nu}}{\mathcal B}{_{23\rho}}-1-\leftidx{_{\nu\nu}}{\mathcal B}{_{23}}-\mathcal B_{23\rho\rho}-\leftidx{_{\nu\nu}}{\mathcal B}{_{23}}\;\mathcal B_{23\rho\rho} \right)=\\=&\;\mathcal B\rvert_1\iiint\Big( \leftidx{_{\mu}}{\mathcal B}{_{12}}\;\leftidx{_{\mu}}{\mathcal B}{_{13}}\;\mathcal B_{23}\;\leftidx{_{\nu}}{\mathcal B}{_{23\rho}}\;\leftidx{_{\nu}}{\mathcal B}{_{23\rho}}-\leftidx{_{\mu}}{\mathcal B}{_{12}}\;\leftidx{_{\mu}}{\mathcal B}{_{13}}\;\mathcal B_{23}-\leftidx{_{\mu}}{\mathcal B}{_{12}}\;\leftidx{_{\mu}}{\mathcal B}{_{13}}\;\mathcal B_{23}\;\leftidx{_{\nu\nu}}{\mathcal B}{_{23}}+\\&-\leftidx{_{\mu}}{\mathcal B}{_{12}}\;\leftidx{_{\mu}}{\mathcal B}{_{13}}\;\mathcal B_{23}\;\mathcal B_{23\rho\rho}-\leftidx{_{\mu}}{\mathcal B}{_{12}}\;\leftidx{_{\mu}}{\mathcal B}{_{13}}\;\mathcal B_{23}\;\leftidx{_{\nu\nu}}{\mathcal B}{_{23}}\;\mathcal B_{23\rho\rho} \Big)=\\
=&\;\mathcal B\rvert_1\iiint\Big[ -\leftidx{_{\nu}}{\mathcal B}{_{23}}\;\leftidx{_{\mu}}{\mathcal B}{_{12}}\left( \leftidx{_{\mu}}{\mathcal B}{_{13}}\;\mathcal B_{23}\;\leftidx{_{\nu}}{\mathcal B}{_{23\rho}} \right)_{\rho}-\leftidx{_{\mu}}{\mathcal B}{_{12}}\;\leftidx{_{\mu}}{\mathcal B}{_{13}}\;\mathcal B_{23}\left( 1+2\leftidx{}{\mathcal B}{_{23\rho\rho}} \right)+\\&+\leftidx{_{\nu}}{\mathcal B}{_{23}}\;\leftidx{_{\mu}}{\mathcal B}{_{13}}\left( \leftidx{_{\mu}}{\mathcal B}{_{12}}\;\mathcal B_{23}\;\mathcal B_{23\rho\rho} \right)_{\nu} \Big]=\\
=&\;\mathcal B\rvert_1\iiint\Big[ -\leftidx{_{\nu}}{\mathcal B}{_{23}}\;\leftidx{_{\mu}}{\mathcal B}{_{12}}\;\leftidx{_{\mu}}{\mathcal B}{_{13\rho}}\;\mathcal B_{23}\;\leftidx{_{\nu}}{\mathcal B}{_{23\rho}}-\leftidx{_{\nu}}{\mathcal B}{_{23}}\;\leftidx{_{\mu}}{\mathcal B}{_{12}}\;\leftidx{_{\mu}}{\mathcal B}{_{13}}\;\mathcal B_{23\rho}\;\leftidx{_{\nu}}{\mathcal B}{_{23\rho}}+\\&-\cancel{\leftidx{_{\nu}}{\mathcal B}{_{23}}\;\leftidx{_{\mu}}{\mathcal B}{_{12}}\;\leftidx{_{\mu}}{\mathcal B}{_{13}}\;\mathcal B_{23}\;\leftidx{_{\nu}}{\mathcal B}{_{23\rho\rho}}}-\leftidx{_{\mu}}{\mathcal B}{_{12}}\;\leftidx{_{\mu}}{\mathcal B}{_{13}}\;\mathcal B_{23}\left( 1+2\leftidx{}{\mathcal B}{_{23\rho\rho}} \right)+\leftidx{_{\nu}}{\mathcal B}{_{23}}\;\leftidx{_{\mu}}{\mathcal B}{_{12\nu}}\;\leftidx{_{\mu}}{\mathcal B}{_{13}}\;\mathcal B_{23}\;\mathcal B_{23\rho\rho}+\\&+\leftidx{_{\nu}}{\mathcal B}{_{23}}\;\leftidx{_{\mu}}{\mathcal B}{_{12}}\;\leftidx{_{\mu}}{\mathcal B}{_{13}}\;\leftidx{_{\nu}}{\mathcal B}{_{23}}\;\mathcal B_{23\rho\rho}\;+\cancel{\leftidx{_{\nu}}{\mathcal B}{_{23}}\;\leftidx{_{\mu}}{\mathcal B}{_{12}}\;\leftidx{_{\mu}}{\mathcal B}{_{13}}\;\mathcal B_{23}\;\leftidx{_{\nu}}{\mathcal B}{_{23\rho\rho}}}   \Big]=\\
=&\;\mathcal B\rvert_1\iiint\Big[ \underbrace{\leftidx{_{\nu}}{\mathcal B}{_{23}}\;\leftidx{_{\mu\mu}}{\mathcal B}{_{12}}\;\mathcal B_{13\rho}\;\mathcal B_{23}\;\leftidx{_{\nu}}{\mathcal B}{_{23\rho}}}_{\equiv\;\mathcal I}\underbrace{-\leftidx{_{\nu}}{\mathcal B}{_{23}}\;\leftidx{_{\mu}}{\mathcal B}{_{12}}\;\leftidx{_{\mu}}{\mathcal B}{_{13}}\;\mathcal B_{23\rho}\;\leftidx{_{\nu}}{\mathcal B}{_{23\rho}}}_{\equiv \; \mathcal J}+\\&-\leftidx{_{\mu}}{\mathcal B}{_{12}}\;\leftidx{_{\mu}}{\mathcal B}{_{13}}\;\mathcal B_{23}\left( 1+2\leftidx{}{\mathcal B}{_{23\rho\rho}} \right)-\leftidx{_{\nu}}{\mathcal B}{_{23}}\;\mathcal B_{12\nu}\;\leftidx{_{\mu\mu}}{\mathcal B}{_{13}}\;\mathcal B_{23}\;\mathcal B_{23\rho\rho}+\leftidx{_{\nu}}{\mathcal B}{_{23}}\;\leftidx{_{\mu}}{\mathcal B}{_{12}}\;\leftidx{_{\mu}}{\mathcal B}{_{13}}\;\leftidx{_{\nu}}{\mathcal B}{_{23}}\;\mathcal B_{23\rho\rho}   \Big],
\end{split}
\label{81}
\end{equation}
with
\begin{equation}
\begin{split}
\mathcal I:=&\;\frac{1}{2}\left( \leftidx{_{\nu}}{\mathcal B}{_{23}}{}^2 \right)_{\rho}\;\leftidx{_{\mu\mu}}{\mathcal B}{_{12}}\;\mathcal B_{13\rho}\;\mathcal B_{23}\simeq-\frac{1}{2}\leftidx{_{\nu}}{\mathcal B}{_{23}}{}^2\;\leftidx{_{\mu\mu}}{\mathcal B}{_{12}}\;\mathcal B_{13\rho\rho}\;\mathcal B_{23}-\frac{1}{2}\leftidx{_{\nu}}{\mathcal B}{_{23}}{}^2\;\leftidx{_{\mu\mu}}{\mathcal B}{_{12}}\;\mathcal B_{13\rho}\;\mathcal B_{23\rho}\\
\mathcal J:=&\;\leftidx{_{\nu}}{\mathcal B}{_{23}}\;\leftidx{_{\mu\mu}}{\mathcal B}{_{12}}\;\mathcal B_{13}\;\mathcal B_{23\rho}\;\leftidx{_{\nu}}{\mathcal B}{_{23\rho}}=\frac{1}{2}\left( \leftidx{_{\nu}}{\mathcal B}{_{23}}{}^2 \right)_{\rho}\;\leftidx{_{\mu\mu}}{\mathcal B}{_{12}}\;\mathcal B_{13}\;\mathcal B_{23\rho}\simeq-\frac{1}{2}\leftidx{_{\nu}}{\mathcal B}{_{23}}{}^2\;\leftidx{_{\mu\mu}}{\mathcal B}{_{12}}\;\mathcal B_{13\rho}\;\mathcal B_{23\rho}+\\&-\frac{1}{2}\leftidx{_{\nu}}{\mathcal B}{_{23}}{}^2\;\leftidx{_{\mu\mu}}{\mathcal B}{_{12}}\;\mathcal B_{13}\;\mathcal B_{23\rho\rho},
\end{split}
\label{82}
\end{equation}
where the symbol ``$\simeq$'' means equal up to an irrelevant  integration by parts. Hence
\begin{equation*}
\begin{split}
&\mathcal B\rvert_1\iiint\Big[ \mathcal I +\mathcal J-\leftidx{_{\mu}}{\mathcal B}{_{12}}\;\leftidx{_{\mu}}{\mathcal B}{_{13}}\;\mathcal B_{23}\left( 1+2\leftidx{}{\mathcal B}{_{23\rho\rho}} \right)-\leftidx{_{\nu}}{\mathcal B}{_{23}}\;\mathcal B_{12\nu}\;\leftidx{_{\mu\mu}}{\mathcal B}{_{13}}\;\mathcal B_{23}\;\mathcal B_{23\rho\rho}+\\&+\leftidx{_{\nu}}{\mathcal B}{_{23}}\;\leftidx{_{\mu}}{\mathcal B}{_{12}}\;\leftidx{_{\mu}}{\mathcal B}{_{13}}\;\leftidx{_{\nu}}{\mathcal B}{_{23}}\;\mathcal B_{23\rho\rho}   \Big]=\\
=&\;\mathcal B\rvert_1\iiint\Big[ -\frac{1}{2}\leftidx{_{\nu}}{\mathcal B}{_{23}}{}^2\;\leftidx{_{\mu\mu}}{\mathcal B}{_{12}}\;\mathcal B_{13\rho\rho}\;\mathcal B_{23}-\frac{1}{2}\leftidx{_{\nu}}{\mathcal B}{_{23}}{}^2\;\leftidx{_{\mu\mu}}{\mathcal B}{_{12}}\;\mathcal B_{13\rho}\;\mathcal B_{23\rho} +\\&-\frac{1}{2}\leftidx{_{\nu}}{\mathcal B}{_{23}}{}^2\;\leftidx{_{\mu\mu}}{\mathcal B}{_{12}}\;\mathcal B_{13\rho}\;\mathcal B_{23\rho}-\frac{1}{2}\leftidx{_{\nu}}{\mathcal B}{_{23}}{}^2\;\leftidx{_{\mu\mu}}{\mathcal B}{_{12}}\;\mathcal B_{13}\;\mathcal B_{23\rho\rho}-\leftidx{_{\mu}}{\mathcal B}{_{12}}\;\leftidx{_{\mu}}{\mathcal B}{_{13}}\;\mathcal B_{23}\left( 1+2\leftidx{}{\mathcal B}{_{23\rho\rho}} \right)+\\&-\leftidx{_{\nu}}{\mathcal B}{_{23}}\;\mathcal B_{12\nu}\;\leftidx{_{\mu\mu}}{\mathcal B}{_{13}}\;\mathcal B_{23}\;\mathcal B_{23\rho\rho}+\leftidx{_{\nu}}{\mathcal B}{_{23}}\;\leftidx{_{\mu}}{\mathcal B}{_{12}}\;\leftidx{_{\mu}}{\mathcal B}{_{13}}\;\leftidx{_{\nu}}{\mathcal B}{_{23}}\;\mathcal B_{23\rho\rho}   \Big]\xrightarrow[]{D\to0}\\
&\xrightarrow[]{D\to0}\mathcal B\rvert_1\int\mathrm d\tau_1\int\mathrm d\tau_2\int\mathrm d\tau_3\;\Big[ -\frac{1}{2}\leftidx{^{\bullet\hspace{-0.1cm}}}{\mathcal B}{_{23}}{}^2\;\leftidx{^{\bullet\bullet\hspace{-0.1cm}}}{\mathcal B}{_{12}}\;\leftidx{}{\mathcal B}{^{\bullet\bullet}_{13}}\;\mathcal B_{23}-\leftidx{^{\bullet\hspace{-0.1cm}}}{\mathcal B}{_{23}}{}^2\;\leftidx{^{\bullet\bullet\hspace{-0.1cm}}}{\mathcal B}{_{12}}\;\mathcal B^{\bullet}_{13}\;\mathcal B^{\bullet}_{23}+
\end{split}
\end{equation*}
\begin{equation}
\begin{split}
&-\frac{1}{2}\leftidx{^{\bullet\hspace{-0.1cm}}}{\mathcal B}{_{23}}{}^2\;\leftidx{^{\bullet\bullet\hspace{-0.1cm}}}{\mathcal B}{_{12}}\;\leftidx{}{\mathcal B}{_{13}}\;\mathcal B_{23}^{\bullet\bullet}-\leftidx{^{\bullet\hspace{-0.1cm}}}{\mathcal B}{_{12}}\;\leftidx{^{\bullet\hspace{-0.1cm}}}{\mathcal B}{_{13}}\;\mathcal B_{23}\left( 1+2\mathcal B^{\bullet\bullet}_{23} \right)-\leftidx{^{\bullet\hspace{-0.1cm}}}{\mathcal B}{_{23}}\;\mathcal B_{12}^{\bullet}\;\leftidx{^{\bullet\bullet\hspace{-0.1cm}}}{\mathcal B}{_{13}}\;\mathcal B_{23}\;\mathcal B_{23}^{\bullet\bullet}+\\&+\leftidx{^{\bullet\hspace{-0.1cm}}}{\mathcal B}{_{23}}\;\leftidx{^{\bullet\hspace{-0.1cm}}}{\mathcal B}{_{12}}\;\leftidx{^{\bullet\hspace{-0.1cm}}}{\mathcal B}{_{13}}\;\leftidx{^{\bullet\hspace{-0.1cm}}}{\mathcal B}{_{23}}\;\mathcal B_{23}^{\bullet\bullet} \Big]=\\=&\;\frac{1}{1440}.
\end{split}
\label{83}
\end{equation}
In the first line of \eqref{81} we have $\leftidx{^{\bullet\hspace{-0.1cm}}}{\mathcal B_{23}}{^{\bullet}}{}^2-\Delta_{\text{gh},23}{}^2$, which needs to be regularized. To do that we adopt the worldline dimensional regularization scheme studied in~\cite{Bastianelli:2000nm, Bastianelli:2006rx}. We introduce $D$ arbitrary dimensions for each worldline integral, {\it i.e.} we extend the worldline time variable to a $(D+1)$-dimensional vector $t\defeq(\tau,t^1,\ldots,t^D)$ along with its derivatives
\begin{equation}
\begin{split}
\leftidx{_{\mu}}{\mathcal B}{}(t_1,t_2)\equiv& \frac{\partial}{\partial t_1{}^{\mu}}\mathcal B(t_1,t_2)\\
\mathcal B_{\nu}(t_1,t_2)\equiv&\frac{\partial}{\partial t_2{}^{\nu}}\mathcal B(t_1,t_2)
\end{split}
\end{equation}
and we integrate over the $(D+1)$-dimensional space. Now, by means of successive integrations by parts we remove ambiguous expressions, neglecting all boundary terms because of momentum conservation in the new $D$ dimensions and periodicity of the propagators in the original interval. We proceed until the final expression is written in a manner that can be unambiguously computed removing the additional dimensions.  

%

\end{document}